\documentclass[a4paper,fleqn,usenatbib]{mnras}
\usepackage{enumerate}
\usepackage[T1]{fontenc}
\usepackage{ae,aecompl}
\usepackage{indentfirst}
\usepackage{graphicx}
\usepackage{amsmath}
\usepackage{amssymb}
\usepackage{color}
\usepackage{mathptmx}
\usepackage{epsfig}
\usepackage{wasysym}
\usepackage{xfrac}
\usepackage{pifont}
\usepackage[table,xcdraw]{xcolor}
\usepackage{booktabs}
\usepackage{multirow}
\usepackage{comment}
\usepackage{longtable}
\usepackage[normalem]{ulem}
\usepackage{subcaption}
\title[Dwarf galaxy structure]{The structural properties of nearby dwarf galaxies in low density environments -- size, surface brightness and colour gradients}

\author[I. Lazar et al.]{
I. Lazar\thanks{E-mail: i.lazar@herts.ac.uk},$^{1}$ S. Kaviraj$^{1}$, A. E. Watkins,$^{1}$ G. Martin,$^{2}$ B. Bichang'a$^{1}$ and R. A. Jackson$^{3}$
\\
$^{1}$Centre for Astrophysics Research, University of Hertfordshire, College Lane, Hatfield AL10 9AB, UK\\
$^{2}$School of Physics and Astronomy, University of Nottingham, University Park, Nottingham NG7 2RD, UK\\
$^{3}$Department of Physics and Astronomy, University of Victoria, Victoria, BC, Canada V8P 5C2\\
}
\pubyear{2024}

\begin{document}
\label{firstpage}
\pagerange{\pageref{firstpage}--\pageref{lastpage}}
\maketitle

\begin{abstract}
We use a complete sample of 211 nearby ($z<0.08$) dwarf (10$^{8}$ M$_{\odot}$ < $M_{\rm{\star}}$ < 10$^{9.5}$ M$_{\odot}$) galaxies in low-density environments, to study their structural properties: effective radii ($R_{\rm e }$), effective surface brightnesses ($\langle \mu \rangle_{\rm e}$) and colour gradients. We explore these properties as a function of stellar mass and the three principal dwarf morphological types identified in a companion paper (Lazar et al.) -- early-type galaxies (ETGs), late-type galaxies (LTGs) and featureless systems. The median $R_{\rm e }$ of LTGs and featureless galaxies are factors of $\sim$2 and $\sim$1.2 larger than the ETGs. While the median $\langle \mu \rangle_{\rm e}$ of the ETGs and LTGs is similar, the featureless class is $\sim$1 mag arcsec$^{-2}$ fainter. Although they have similar median $R_{\rm e }$, the featureless and ETG classes differ significantly in their median $\langle \mu \rangle_{\rm e}$, suggesting that their evolution is different and that the featureless galaxies are not a subset of the ETGs. While massive ETGs typically exhibit negative or flat colour gradients, dwarf ETGs generally show positive colour gradients (bluer centres). The growth of ETGs therefore changes from being `outside-in' to `inside-out' as we move from the dwarf to the massive regime. The colour gradients of dwarf and massive LTGs are, however, similar. Around 46 per cent of dwarf ETGs show prominent, visually-identifiable blue cores which extend out to $\sim$1.5 $R_{\rm e}$. Finally, compared to their non-interacting counterparts, interacting dwarfs are larger, bluer at all radii and exhibit similar median $\langle \mu \rangle_{\rm e}$, indicating that interactions typically enhance star formation across the entire galaxy.

\end{abstract}



\section{Introduction}

Dwarf galaxies ($M_\star$ < 10$^{9.5}$ M$_\odot$) dominate the galaxy census in all environments \citep[e.g.][]{Ferguson1994,Baldry2012,Alavi2016}, 
making them important for a complete comprehension of galaxy evolution. The physical characteristics of dwarfs make them particularly useful for probing the physics of galaxy evolution. For example, their shallow potential wells make dwarfs more sensitive probes of key physical processes, such as baryonic feedback, tidal perturbations, and ram pressure than their massive counterparts \citep[e.g.][]{Martin2019,Jackson2021a}. Similarly, the high dark-matter fractions in these systems \citep[e.g.][]{Kirby2015,Jackson2021b,Bruce2023,Jackson2024}, make them good laboratories for probing the nature of dark matter \citep[e.g.][]{Ackermann2015}.

In spite of their utility as cosmological probes, much of our current understanding of dwarf galaxies comes from studies in the very nearby Universe. This is because typical dwarfs are not bright enough to be detectable in the shallow surveys, e.g. the SDSS \citep{Alam2015}, that have dominated the astrophysical landscape over the past few decades \citep[e.g.][]{Lazar2024}. The dwarfs that do appear in these surveys tend to have high star formation rates (SFRs) which make them brighter and detectable in shallow surveys \citep[e.g.][]{Jackson2021a}. However, this also biases them towards objects that tend to be bluer and may skew the morphological mix towards late-type systems \citep[e.g.][]{Lazar2024}. Unbiased studies of the dwarf population outside our local neighbourhood requires surveys that are both deep and wide (as is the case here).   

A rich literature exists on key structural properties in massive galaxies, such as galaxy size (typically parametrised by the half-light or `effective' radius, $R_{\rm e}$), the effective surface-brightness ($\langle \mu \rangle_{\rm e}$, defined here as the mean surface brightness within the effective radius) and colour profiles and gradients. In the massive-galaxy regime, these structural parameters tend to change both as a function of stellar mass and morphology. For example, around $M_\star$ $\sim$ 10$^{10}$ M$_\odot$, massive early-type galaxies (ETGs) have $R_{\rm e}$ values that are around a factor of 2 smaller and $\langle \mu \rangle_{\rm e}$ values that are around a factor of 2.5 brighter than late-type galaxies (LTGs) \citep[e.g.][]{Shen2003}. However, at higher stellar masses ($M_{\rm \star}$ $\sim$ 10$^{11}$ M$_\odot$) massive ETGs and LTGs exhibit similar $R_{\rm e}$ and $\langle \mu \rangle_{\rm e}$ \citep[e.g.][]{Shen2003,Graham2008}, while beyond this value ETGs become larger than their LTG counterparts. The colour gradients in massive ETGs are generally flat or negative i.e. massive ETGs are typically either red throughout or redder in their centres \citep[e.g.][]{DePropris2005,LaBarbera2012,Ferreras2017}, with the bluer outer regions attributed to late-epoch satellite accretion. Massive LTGs, on the other hand, generally have negative colour gradients in the inner regions and flat or positive gradients in the outskirts \citep[e.g.][]{DSouza2014,Tortora2010,Liao2023}. The progressive reddening in the outskirts is attributed to non-axisymmetric features, such as spirals or bars and/or a star formation cut off due to a surface density threshold \citep[e.g.][]{Roskar2008,Martin2001}.

While dwarf galaxies that are detected in shallow surveys, but which reside outside the local neighbourhood, are likely to show the biases outlined above, relatively unbiased explorations of dwarfs are possible in the Local Group \citep[e.g.][]{McConnachie2012} or in the very nearby Universe via surveys such as MATLAS \citep[e.g.][]{Duc2015}, NGVS \citep[][]{Ferrarese2012,Ferrarese2020} and FDS \citep[e.g.][]{Venhola2017,Venhola2019} which identify dwarfs either around nearby massive galaxies or in nearby groups and clusters. By construction, such studies offer insights into dwarf galaxy evolution in relatively high-density environments.

In such dense environments, past work has found four main morphological classes: dwarf ellipticals, which are systems with central light concentrations and smooth light distributions like those found in the massive-galaxy regime \citep[e.g.][]{Graham2003a,Janz2017}, dwarf `spheroidals', which are diffuse low-surface-brightness systems which lack a central light concentration \citep[e.g.][]{Irwin1990,Wilkinson2006,Simon2007,Koda2015}, dwarf spirals, which are rotationally-supported systems with spiral structure \citep[e.g.][]{Schombert1995,Graham2003} and dwarf irregulars which appear chaotic in their structural appearance and are thought to host significant amounts of gas \citep[e.g.][]{Gallagher1984,Hunter2024}. 

Several studies \citep[e.g.][]{McConnachie2012,vanDokkum2015,Eigenthaler2018,Lim2020,Ferrarese2020,Poulain2021} have compared the structural properties of these types. For example, at a stellar mass of $M_{\rm \star}$ $\sim$ 10$^{8}$ M$_\odot$, the $R_{\rm e}$ values of dwarf ellipticals and irregulars are similar ($\sim$ 1 kpc in the $g$-band), with the dwarf spheroidals typically larger by around a factor of 2. At such stellar masses the typical $\langle \mu \rangle_{\rm e}$ of dwarf ellipticals and irregulars (in the $g$-band) is $\sim$ 24 mag arcsec$^{-2}$, while dwarf spheroidals are a factor of $\sim$4 fainter. At lower stellar masses, e.g. $M_{\rm \star}$ $\sim$ 10$^{7}$ M$_\odot$, all dwarf morphological types show a similar large scatter around effective radii and effective surface brightness values of $R_{\rm e,g}$ $\sim$ 0.7 kpc and $\langle \mu \rangle_{\rm e,g}$ $\sim$ 25 mag arcsec$^{-2}$ respectively. The colour gradients in dwarfs that reside in high-density environments are found to mostly vary from flat or negative at the upper end of the stellar masses spanned by dwarfs ($M_{\rm \star}$ $\sim$ 10$^{9.5}$ M$_\odot$), to flat or positive at lower stellar masses \citep[$M_{\rm \star}$ $\sim$ 10$^{8}$ M$_\odot$, e.g.][]{Vader1988,Bremnes2000,Jansen2000,Parodi2002,Tortora2010,denBrok2011}. Dwarf irregulars, in particular, generally have positive gradients \citep[e.g.][]{Parodi2002}.

While the studies described above have shaped our understanding of dwarfs in nearby high-density regions, much less is known about the bulk of the dwarf population that lives in low-density environments. To address this, we have constructed, in a recent paper \citep{Lazar2024}, a mass-complete, unbiased sample of 257 dwarf (10$^{8}$ M$_{\odot}$ < \textit{M}$_{\rm{\star}}$ < 10$^{9.5}$ M$_{\odot}$) galaxies at $z<0.08$ in the COSMOS field which, at these redshifts hosts galaxies in groups and the field i.e. low-density environments. Visual inspection of ultra-deep optical images of these dwarfs from the Hyper Suprime-Cam reveals three principal morphological classes in dwarfs that inhabit low-density environments: ETGs, i.e. elliptical and S0 systems, LTGs, which show evidence of disks and `featureless' dwarfs which show neither the central light concentration seen in ETGs nor any spiral structure that typifies LTGs. 43, 45 and 10 per cent of dwarfs correspond to the ETG, LTG and featureless class, while a small fraction (2 per cent) are morphologically irregular. The featureless class is akin to the dwarf spheroidals in high-density regions. However, \citet{Lazar2024} label them as featureless rather than spheroidal because, at least in the field, the featureless systems are not produced by cluster-specific processes that are thought to give rise to the dwarf spheroidals in high-density environments. Indeed, as we show later in this study, the featureless dwarfs deviate strongly from the dwarf ETGs in their structural properties, which suggests that the featureless galaxies are not a subset of the ETG population. This is analogous to the findings of \citet[][]{Kormendy2012}, who noted significant structural differences between dwarf spheroidal and dwarf elliptical galaxies in high density environments. It is worth noting that, while the dwarf ETGs and LTGs are akin to the well-established morphologies seen in the massive-galaxy regime \citep[e.g.][]{Hubble1936,Lintott2011,Kaviraj2014}, the featureless class is essentially missing in the massive-galaxy regime. 

\citet{Lazar2024} have used this sample to study the frequency of dwarfs in each morphological class and their key properties e.g. colours, local environments, incidence of interactions and the extent to which the visually classified morphologies can be separated using standard morphological parameters. This paper is a companion study, which focuses on exploring the structural properties of the Lazar et al. dwarfs, as a function of their stellar mass and morphology. In particular, we probe typical structural parameters that underpin similar work in the massive-galaxy regime: size (parametrised by $R_{\rm e}$), $\langle \mu \rangle_{\rm e}$ and colour profiles and gradients. The overall aim of this study, when combined with \citet{Lazar2024}, is to establish a low-redshift benchmark for the morphological and structural properties of dwarf galaxies in low-density environments. This is particularly desirable, given the imminent arrival of data from deep-wide surveys like the Legacy Survey of Space and Time \citep[e.g.][]{Ivezic2019,Watkins2024}, which will enable similar analyses over large areas of the sky. 

This paper is organised as follows. In Section \ref{sec:cosmos2020}, we describe the sample of nearby dwarf galaxies, from \citet{Lazar2024}, that underpins this study. In Sections \ref{sec:sizes} and \ref{sec:sb}, we explore the values of $R_{\rm e}$ and $\langle \mu \rangle_{\rm e}$ of our dwarfs, as a function of stellar mass and morphology. In Section \ref{sec:colours_gradients}, we study the colour gradients of our dwarfs as a function of stellar mass and morphology. We summarise our findings in Section \ref{sec:summary}.


\section{A sample of nearby dwarf galaxies}
\label{sec:cosmos2020}

Our study is based on the dwarf galaxy sample constructed by \citet{Lazar2024}. In this section we describe aspects of the creation of this catalogue that are relevant to our study. We direct readers to \citet{Lazar2024} for further details about its construction. The sample is assembled using the Classic version of the COSMOS2020 catalogue \citep{Weaver2022}. This catalogue provides accurate physical parameters (e.g. photometric redshifts, stellar masses and star formation rates) for $\sim$1.7 million sources in the $\sim$2 deg$^2$ COSMOS field \citep{Scoville2007}. The parameter estimation employs deep photometry in 40 broadband filters spanning the UV through to the mid-infrared, from the following instruments: GALEX \citep{Zamojski2007}, MegaCam/CFHT \citep{Sawicki2019}, ACS/HST \citep{Leauthaud2007}, Hyper Suprime-Cam \citep{Aihara2019}, Subaru/Suprime-Cam \citep{Taniguchi2007,Taniguchi2015}, VIRCAM/VISTA \citep{McCracken2012} and IRAC/Spitzer \citep{Ashby2013,Steinhardt2014,Ashby2015,Ashby2018}. Aperture photometry in the optical and infrared filters is extracted using the \textsc{SExtractor} \citep{Bertin1996} and \textsc{IRACLEAN} \cite{Hsieh2012} codes respectively. The physical parameters are then calculated  using the \textsc{LePhare} SED-fitting algorithm \citep{Arnouts2002,Ilbert2006}. The wide wavelength baseline results in photometric redshift accuracies better than $\sim$1 and $\sim$4 per cent for bright ($i<22.5$ mag) and faint ($25<i<27$ mag) galaxies respectively.

\begin{figure*}
 \centering
 \includegraphics[width=0.899\textwidth]{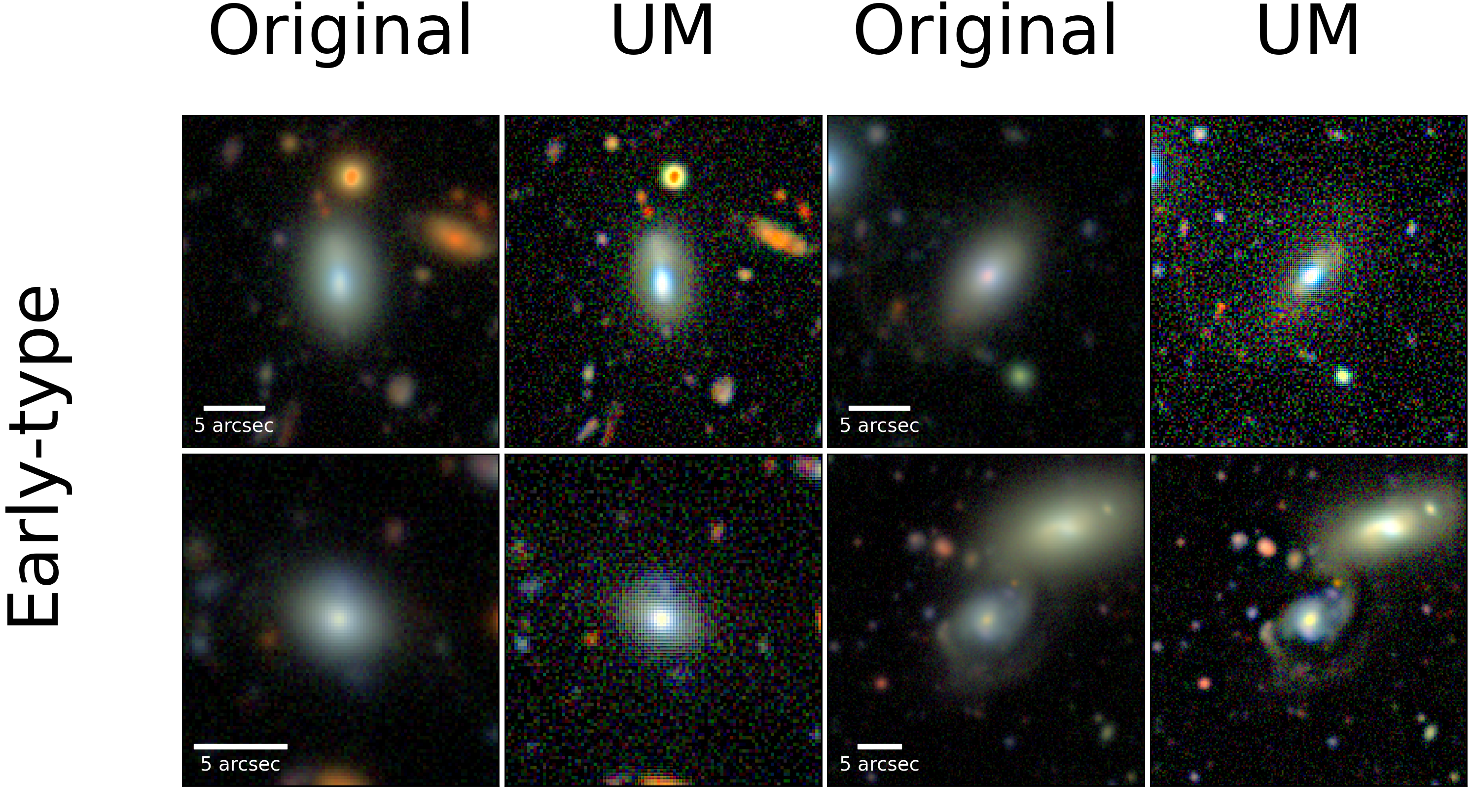}
 \includegraphics[width=0.895\textwidth]{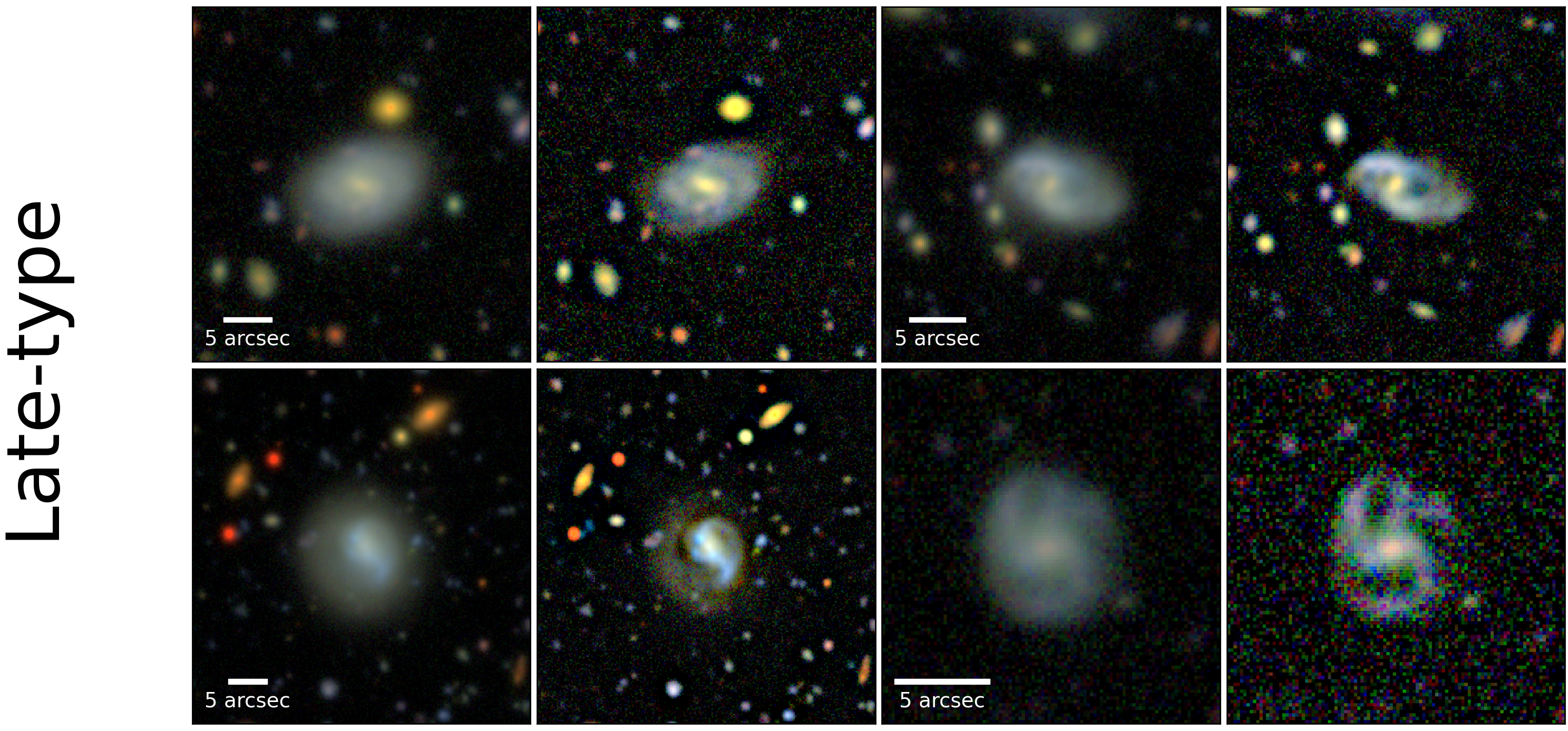}
 \includegraphics[width=0.895\textwidth]{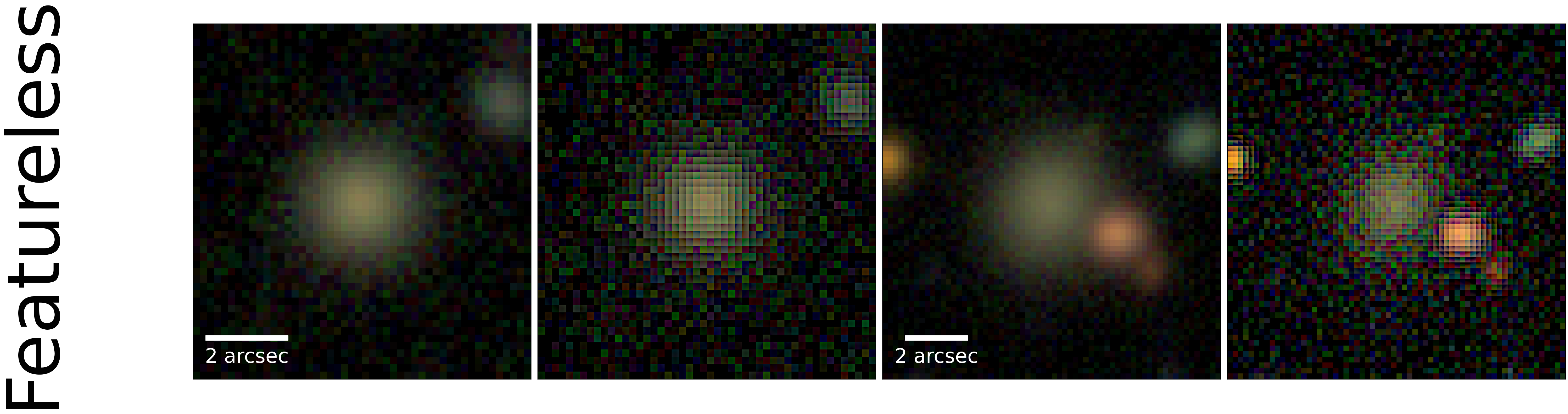}
 \caption{Examples of the three principal morphological classes -- early-type, late-type and featureless galaxies -- found by \citet{Lazar2024} in nearby ($z<0.08$) dwarf galaxies. For each galaxy, we show the original $gri$ colour-composite image and its unsharp-masked counterpart (in the columns labelled `UM'). The Hyper Suprime-Cam images have a point-source depth of $\sim$28 mag, around 5 magnitudes deeper than standard-depth SDSS imaging. Examples of interactions include internal asymmetries (e.g. row 1, cols 1/2 and row 4, cols 1/2), faint tidal features (e.g. row 1, cols 3/4; the feature is visible to the north-east of the galaxy) and tidal bridges in ongoing interactions (e.g. row 2, cols 3/4).}. 
 \label{fig:images}
\end{figure*}

\begin{table*}
\centering
\begin{tabular}{cccccc}
 & \multicolumn{5}{c}{Median $R_{\rm e}$ (arcsec)}                                                     \\ \hline Morphology
 & \multicolumn{1}{|c|}{All galaxies} & \multicolumn{1}{c|}{Red} & \multicolumn{1}{c|}{Blue} & \multicolumn{1}{c|}{Interacting} & Non-interacting  \\ \hline dETG
 & \multicolumn{1}{|c|}{1.18 $_{0.03}^{0.08}$} & \multicolumn{1}{c|}{1.26 $_{0.13}^{0.10}$} & \multicolumn{1}{c|}{1.17 $_{0.03}^{0.05}$} & \multicolumn{1}{c|}{1.65 $_{0.17}^{0.17}$} & 1.14 $_{0.04}^{0.03}$ \rule{0pt}{3ex}    \\ dLTG
 & \multicolumn{1}{|c|}{2.06 $_{0.07}^{0.07}$} & \multicolumn{1}{c|}{2.12 $_{0.11}^{0.13}$} & \multicolumn{1}{c|}{2.01 $_{0.05}^{0.12}$} & \multicolumn{1}{c|}{2.22 $_{0.17}^{0.17}$} & 2.00 $_{0.05}^{0.08}$ \rule{0pt}{3ex}  \\ dF
 & \multicolumn{1}{|c|}{1.37 $_{0.11}^{0.06}$} & \multicolumn{1}{c|}{1.22 $_{0.14}^{0.22}$} & \multicolumn{1}{c|}{1.40 $_{0.07}^{0.21}$} & \multicolumn{1}{c|}{1.50 $_{0.17}^{0.17}$} & 1.33 $_{0.20}^{0.09}$ \rule{0pt}{3ex}  \\  All morphologies
 & \multicolumn{1}{|c|}{1.61 $_{0.07}^{0.04}$} & \multicolumn{1}{c|}{1.40 $_{0.05}^{0.08}$} & \multicolumn{1}{c|}{1.66 $_{0.05}^{0.07}$} & \multicolumn{1}{c|}{1.94 $_{0.11}^{0.11}$} & 1.49 $_{0.05}^{0.08}$ \rule{0pt}{3ex}  \\  \noalign{\vskip 1mm}  \hline
\end{tabular}%
\caption{Median $R_{\rm e}$ in the $i$-band for different dwarf galaxy populations (dETG = dwarf ETG, dLTG = dwarf LTG, and dF = dwarf featureless). The errors on the medians (calculated via bootstrapping) are indicated using superscripts and subscripts. The threshold between the red and blue populations is at rest-frame $(g-i)$ = 0.7 (following \citet[][]{Lazar2024}.}
\label{tab:eff_rad}
\end{table*}

\begin{table*}
\centering
\begin{tabular}{cccccc}
 & \multicolumn{5}{c}{Median $\langle \mu \rangle_{\rm e}$ (mag arcsec$^{-2}$)}                                                     \\ \hline Morphology
 & \multicolumn{1}{|c|}{All galaxies} & \multicolumn{1}{c|}{Red} & \multicolumn{1}{c|}{Blue} & \multicolumn{1}{c|}{Interacting} & Non-interacting  \\ \hline dETG
 & \multicolumn{1}{|c|}{22.9 $_{0.3}^{0.1}$} & \multicolumn{1}{c|}{23.1 $_{0.1}^{0.1}$} & \multicolumn{1}{c|}{22.6 $_{0.1}^{0.3}$} & \multicolumn{1}{c|}{22.8 $_{0.5}^{0.6}$} & 22.9 $_{0.3}^{0.1}$ \rule{0pt}{3ex}    \\ dLTG
 & \multicolumn{1}{|c|}{23.2 $_{0.1}^{0.1}$} & \multicolumn{1}{c|}{23.2 $_{0.4}^{0.3}$} & \multicolumn{1}{c|}{23.2 $_{0.1}^{0.1}$} & \multicolumn{1}{c|}{23.2 $_{0.2}^{0.1}$} & 23.2 $_{0.1}^{0.2}$ \rule{0pt}{3ex}  \\ dF
 & \multicolumn{1}{|c|}{24.0 $_{0.2}^{0.1}$} & \multicolumn{1}{c|}{24.4 $_{0.2}^{0.2}$} & \multicolumn{1}{c|}{23.6 $_{0.1}^{0.2}$} & \multicolumn{1}{c|}{23.8 $_{0.3}^{0.4}$} & 24.0 $_{0.2}^{0.1}$ \rule{0pt}{3ex}  \\  All morphologies
 & \multicolumn{1}{|c|}{23.2 $_{0.1}^{0.1}$} & \multicolumn{1}{c|}{23.4 $_{0.2}^{0.2}$} & \multicolumn{1}{c|}{23.1 $_{0.1}^{0.1}$} & \multicolumn{1}{c|}{23.3 $_{0.2}^{0.1}$} & 23.2 $_{0.1}^{0.1}$ \rule{0pt}{3ex}  \\   \noalign{\vskip 1mm}  \hline
\end{tabular}%
\caption{Median $\langle \mu \rangle_{\rm e}$ in the $i$-band for different galaxy populations (dETG = dwarf ETG, dLTG = dwarf LTG, and dF = dwarf featureless). The errors on the medians (calculated via bootstrapping) are indicated using superscripts and subscripts. The threshold between the red and blue populations is at rest-frame $(g-i)$ = 0.7 (following \citet[][]{Lazar2024}.}
\label{tab:eff_sb}
\end{table*}

The dwarf galaxy sample is then constructed by selecting objects that are classified as galaxies by \textsc{LePhare} (`type' = 0 in the COSMOS2020 catalogue), exhibit an extendedness of 1 in the HSC $griz$ filters (i.e. are classified as galaxies by the HSC pipeline), have stellar masses in the range 10$^{8}$ M$_{\odot}$ < M$_{\rm{\star}}$ < 10$^{9.5}$ M$_{\odot}$, redshifts in the range $z<0.08$ and lie outside masked regions. The final sample contains 257 dwarf galaxies, with median redshift and stellar mass errors of 0.02 and 0.08 dex respectively. As noted in \citet{Lazar2024}, given the depth of the data, this dwarf galaxy sample is mass complete and therefore offers an unbiased statistical sample of galaxies which can be used to study the morphological and structural properties of the nearby dwarf population. 

\subsection{Morphological classification via visual inspection}
The galaxies in this final sample are then visually classified, as described in \citet{Lazar2024}, using optical $gri$ colour-composite images, and their unsharp-masked counterparts, from the HSC-SSP Ultra-deep layer. This layer has a 5$\sigma$ point source depth of $\sim$28 magnitudes, around 5 magnitudes deeper than standard depth SDSS imaging and almost 10 magnitudes deeper than the magnitude limit of the SDSS spectroscopic main galaxy sample. The median HSC seeing is $\sim$0.6 arcseconds, around a factor of 2 better than the SDSS. The visual inspection is used to classify the dwarfs into three principal morphological classes: early-type galaxies (ETGs) i.e. elliptical and S0 systems, late-type galaxies (LTGs) which show evidence of disks and featureless galaxies which show neither the central light concentrations seen in ETGs or any disk structure that typifies LTGs. The inspection is also used to flag dwarfs that show evidence of interactions e.g. internal asymmetries, tidal features and tidal bridges that connect galaxies in ongoing mergers. We direct readers to Section 3 in \citet{Lazar2024} for more details of the morphological classification. Figure \ref{fig:images} shows an abridged version of the images presented in \citet{Lazar2024}, with examples of dwarfs in the different morphological classes. Finally, as noted in \citet{Lazar2024}, the COSMOS2020 galaxy population, in our redshift range of interest, resides preferentially in low-density environments (i.e. galaxies in groups and the field).

\subsection{Masking, PSF correction and construction of surface-brightness profiles}
\label{sec:preprocess}

\begin{figure*}
 \centering
 \includegraphics[width=0.95\textwidth]{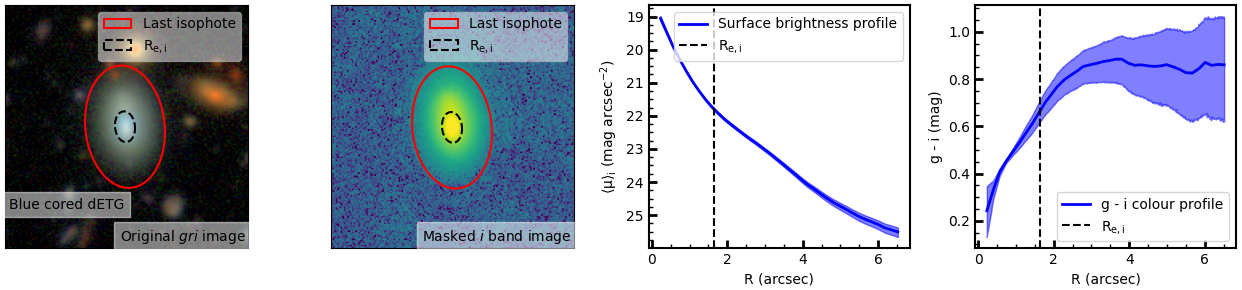}
 \includegraphics[width=0.95\textwidth]{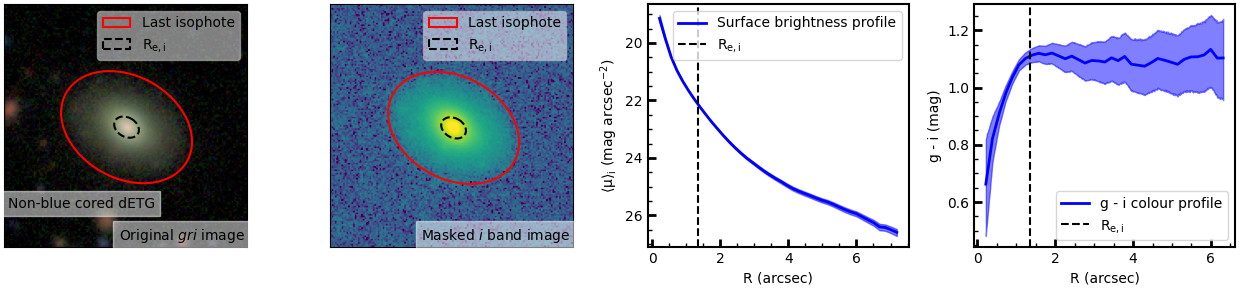}
 \includegraphics[width=0.95\textwidth]{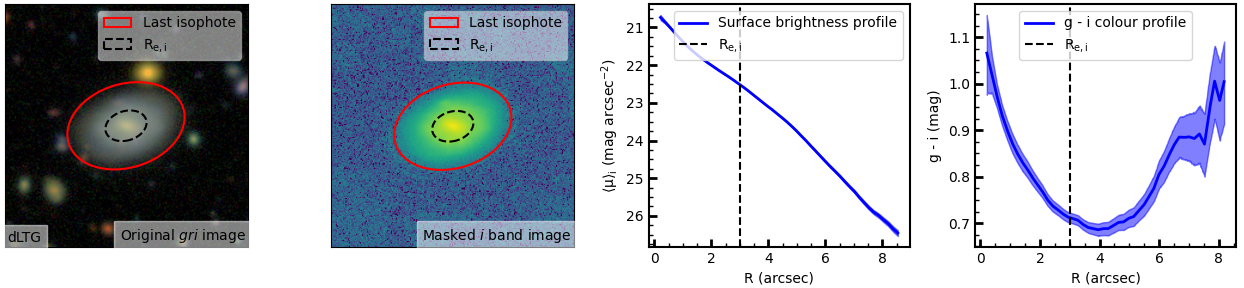}
 \includegraphics[width=0.95\textwidth]{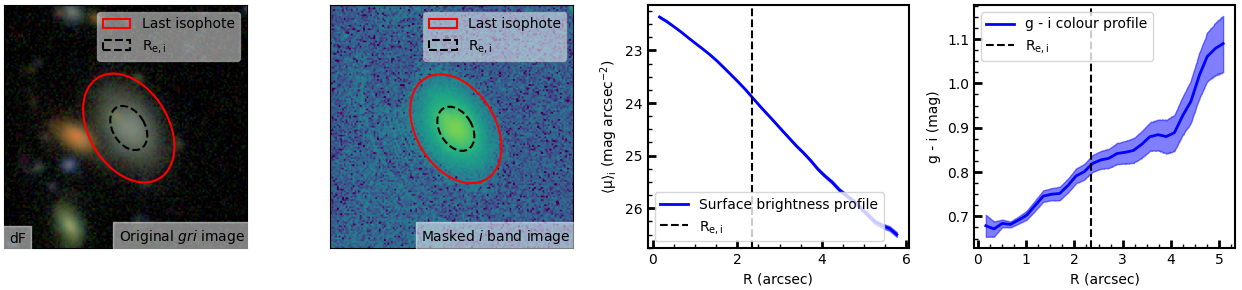}
 \caption{Examples of $i$-band surface brightness and $(g-i)$ colour profiles of dwarfs in different morphological classes (shown in the different rows). The first column (from the left) shows the original $gri$ colour-composite image of the galaxy. The second column shows the masked $i$-band image. The third column presents the $i$-band surface brightness profile ending at the last isophote (shown by the red ellipse in the first 2 columns), which is defined as the surface brightness that corresponds to 2 times the background RMS. The last column presents the $(g-i)$ colour profile ending at the last isophote. The black dashed vertical line (corresponding to the black dashed ellipse in the second column) represents the $i$-band half-light radius (i.e. the effective radius). The morphology of each galaxy is indicated in the lower left corner of each $gri$ galaxy image (dETG = dwarf ETG, dLTG = dwarf LTG, and dF = dwarf featureless). The uncertainties in the surface brightness profiles, shown by the blue shaded regions are calculated following \citet[][]{MartinezLombilla2019}, by taking into account the Poissonian error of each bin and its associated sky RMS value. The uncertainties on the colour profiles, shown using the blue shaded regions, are obtained using standard error propagation.}
 \label{fig:example}
\end{figure*}

In each band, we manually mask flux from interloper sources, such as background galaxies and foreground stars. We use the azimuthal flux interpolation method from \citet[][]{Watkins2022} to reconstruct the missing flux in the masked regions (see \citet[][]{Lazar2024} for further details). Figure \ref{fig:example} presents examples of the original $gri$ colour-composite image (column 1) and the final image produced using this method (column 2), which is then used to calculate structural parameters. The morphology of each galaxy is indicated in the lower left corner of each $gri$ galaxy image (dETG = dwarf ETG, dLTG = dwarf LTG, and dF = dwarf featureless).

We construct surface-brightness profiles for our dwarfs using the \textsc{Python}-based galaxy ellipse fitter \texttt{photutils.isophote.Ellipse} in the \texttt{Photutils} package. This ellipse fitter is based on the IRAF \texttt{ELLIPSE} package \citep[][]{Jedrzejewski1987,Busko1996}. The profiles are corrected for light smearing due to the PSF in the following way, as described in \citet{Watkins2022}:

\begin{itemize}
    \item {\color{black} We first fit 2D single-component S\'ersic profiles \citep{Sersic1963} using the 
     \texttt{astropy} module within the \textsc{Python} package \texttt{statmorph}.} 
    \item We use the resultant S\'ersic index, half-light radius and the magnitude obtained from the fitting to produce a 2D model image for each galaxy using \texttt{GALFIT} \citep[][]{Peng2002,Peng2010}.
    \item We convolve this image with the Hyper Suprime-Cam PSF calculated by \citet[][]{Montes2021}. We use $g$ and $i$ band PSFs for the $g$ and $i$ band galaxy images and the $g$-band PSF for images in other filters (because only $g$ and $i$-band PSFs are available from Montes et al.). Note that, while the results presented in this study only involve the $g$ and $i$ bands, structural parameters are calculated in other bands ($r$ and $z$) and are available to the reader upon request. 
    \item We calculate surface-brightness profiles from the convolved and unconvolved model images using the method described above and calculate the ratio between the two in each isophotal bin. This ratio is a measure of the effect of the PSF on the surface-brightness profile.
    \item Finally, we correct for light smearing due to the PSF by multiplying the initial surface-brightness profile (obtained from the original galaxy image) in each isophotal bin by the ratio calculated in the previous step. 
\end{itemize}

Examples of the final surface-brightness profiles for different dwarf morphologies are presented in column 3 of Figure \ref{fig:example}. The corresponding $(g-i)$ colour profiles are presented in column 4. We also calculate curves of growth using \texttt{photutils.isophote.Ellipse} and apply the same PSF correction procedure to these profiles. Using the corrected curves of growth, we then calculate $R_{\rm e}$ (studied in Section \ref{sec:sizes}) and $\langle \mu \rangle_{\rm e}$ (studied in Section \ref{sec:sb}) for each dwarf galaxy.

Finally, we note that before the PSF correction procedure is applied, we eliminate, from our final sample, galaxies with a S\'ersic index higher than 4 and effective radii smaller than the PSF (i.e. 3 pixels). This is because such objects are either close to being unresolved or exhibit sharply peaked surface brightness profiles which hinder the performance of \texttt{GALFIT} to produce images of the convolved and unconvolved models. Our final sample of objects comprises 211 dwarf galaxies, out of the original sample of 257 in \citet{Lazar2024}.

\section{Size}
\label{sec:sizes}

\begin{figure*}
 \centering
 \includegraphics[width=0.44\textwidth]{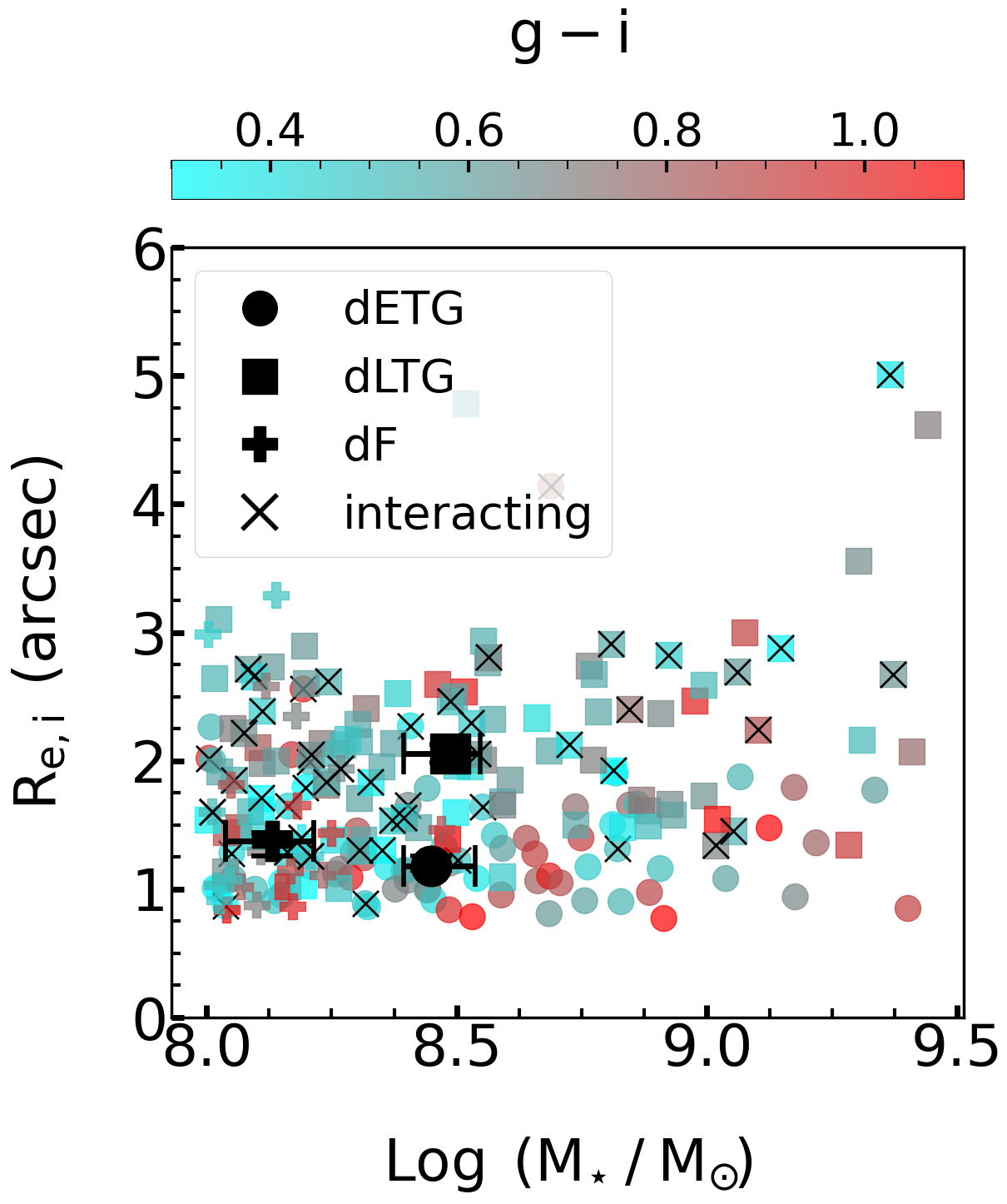}\hspace{0.2in}
  \includegraphics[width=0.499\textwidth]{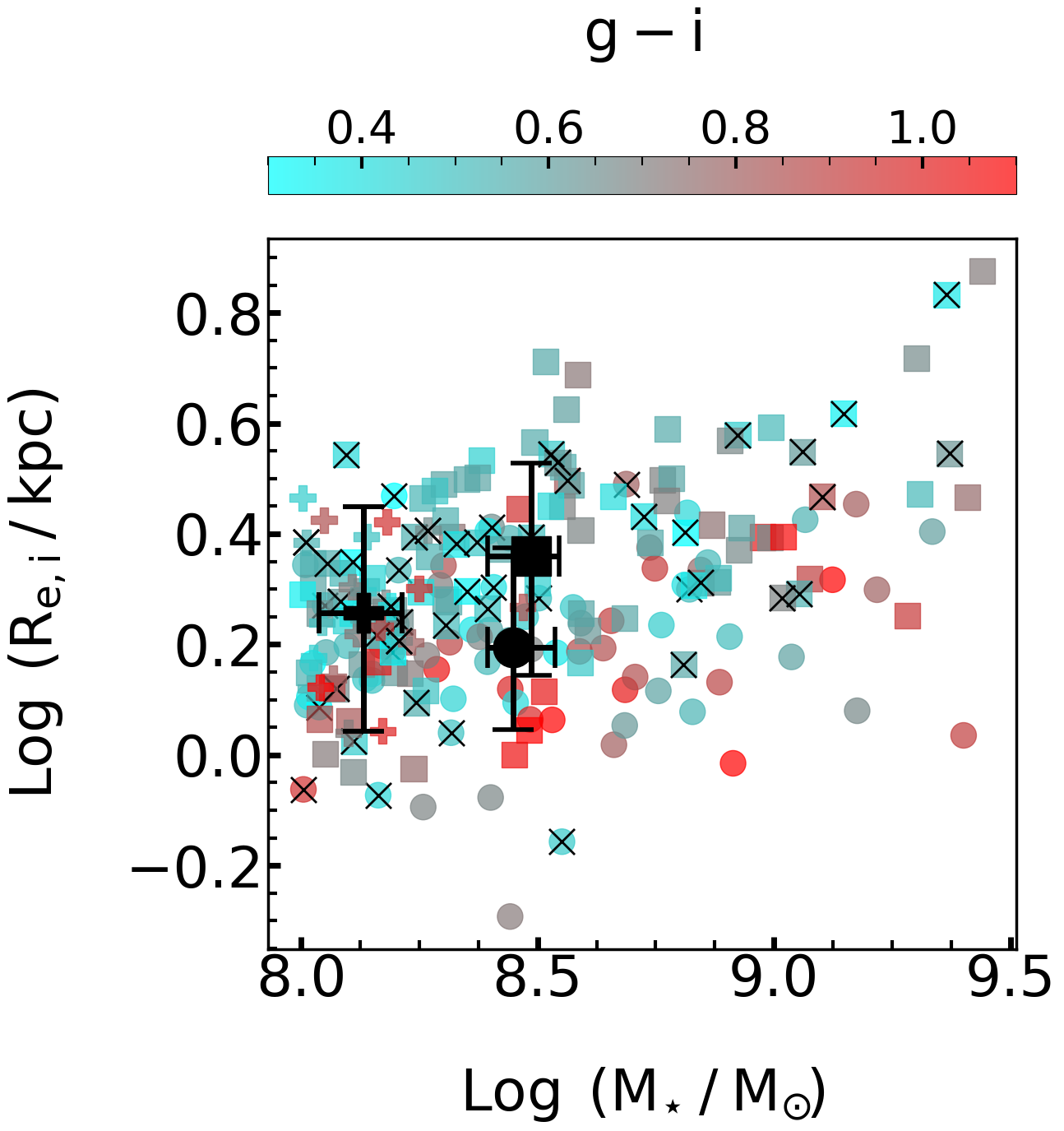}
 \caption{Effective radius in arcseconds (left) and in kpc (right) vs stellar mass for our dwarf galaxies. Different morphological classes are shown using different symbols, while galaxies are colour-coded using their rest-frame $(g-i)$ colour. Galaxies that are interacting are shown using crosses. The median values of $R_{\rm e}$ and stellar mass (and their uncertainties) for each morphological class are shown using the black symbols. Since we use photometric redshifts in this study, the physical sizes (in kpc) incur large errors when the uncertainties in the photometric redshifts are propagated through the conversion from angular to physical effective radii.}
 \label{fig:reff_mass}
\end{figure*}

We begin in Figure \ref{fig:reff_mass} by presenting $R_{\rm e}$ as a function of stellar mass for our dwarf population. Different morphological classes are shown using different symbols, while galaxies are colour-coded using their rest-frame $(g-i)$ colour. Interacting systems are indicated using crosses. In our redshift and stellar mass ranges of interest, the galaxy population is bimodal around $(g-i)$ $\sim$ 0.7 \citep[as shown in][]{Lazar2024}. Red and blue galaxies are defined as those with rest-frame $(g-i)$ greater and less than 0.7 respectively. Table \ref{tab:eff_rad} summarises the median values of the effective radius for different galaxy populations. Note that, since we use photometric redshifts in this study, the physical sizes (in kpc) incur large errors when the uncertainties in the photometric redshifts are propagated through the conversion from angular to physical size (as indicated by the large y-axis error-bars in the right-hand panel of Figure \ref{fig:reff_mass}). Therefore, we largely base our analysis in this section on the angular sizes (left-hand panel of this figure) which, in turn restricts us to studying the \textit{relative} sizes between different morphological classes (rather than physical sizes). However, we also show the physical sizes for completeness in the right-hand panel of this figure. 

\begin{figure}
 \centering
 \includegraphics[width=0.45\textwidth]{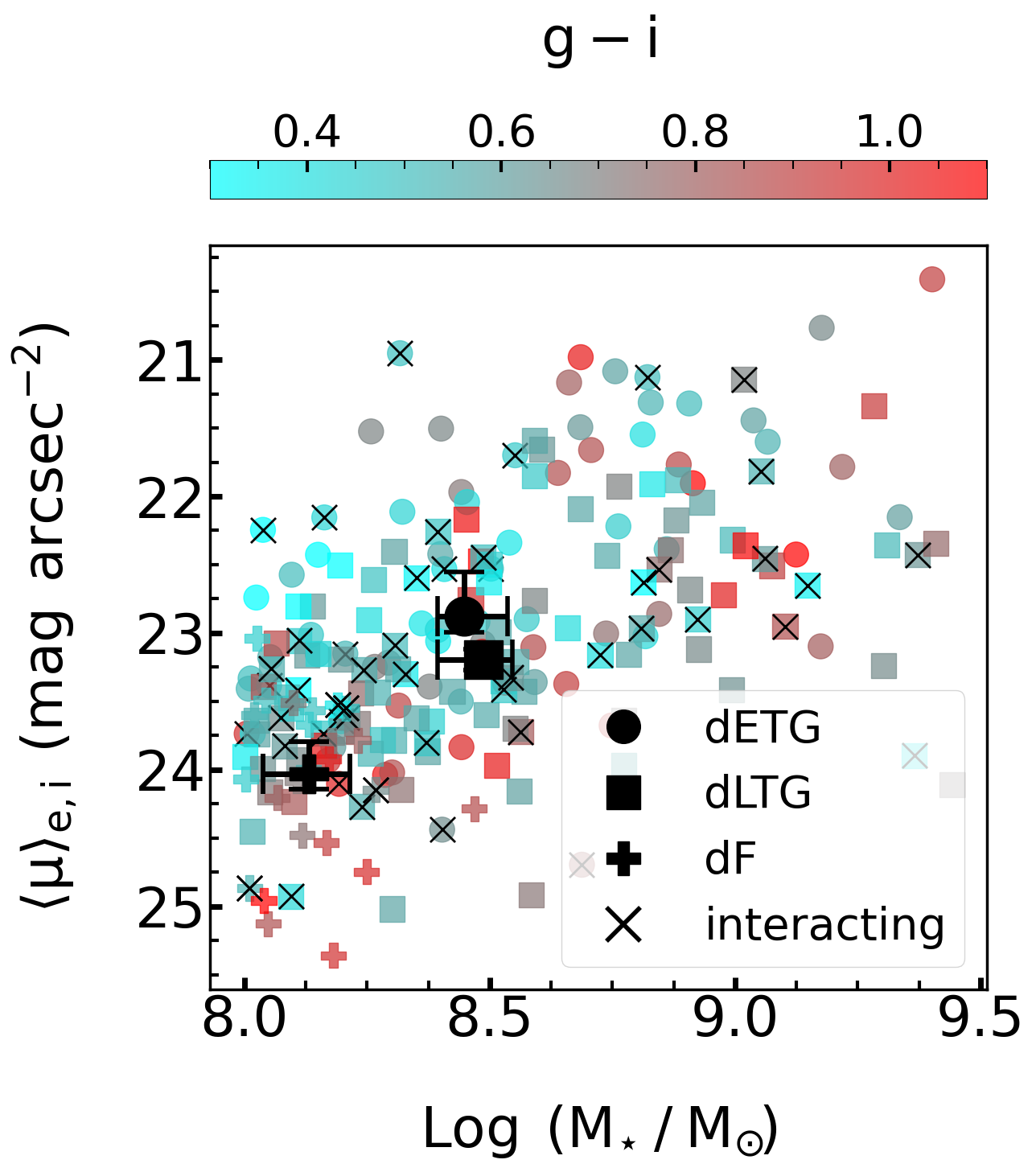}
 \caption{Effective surface brightness vs stellar mass for our dwarf galaxies. Different morphological classes are shown using different symbols, while galaxies are colour-coded using their rest-frame $(g-i)$ colour. Galaxies that are interacting are shown using crosses. The median values of $\langle \mu \rangle_{\rm e}$ and stellar mass (and their uncertainties) for each morphological class are shown using the black symbols.}
 \label{fig:mu_mass}
\end{figure}

Figure \ref{fig:reff_mass} and Table \ref{tab:eff_rad} together indicate that ETGs and LTGs are well separated in $R_{\rm e}$, with the median $R_{\rm e}$ of ETGs being around a factor of 2 smaller than that of LTGs. Note that the two morphological types show similar values of median stellar mass, so this trend is not driven by the ETGs and LTGs having significantly different stellar mass distributions (which can be important since size, to a certain extent, scales with stellar mass). The featureless dwarfs lie in between the ETGs and LTGs, with a median $R_{\rm e}$ that is around a factor of 1.2 larger than their ETG counterparts (in spite of having a median stellar mass than is around 0.5 dex smaller). 

The median values of $R_{\rm e}$ of red (i.e. quenched) and blue (i.e. star-forming) galaxies do not show significant differences within each morphological class. When all morphological classes are considered together, red  galaxies show a weak trend of being more compact than blue galaxies. Table \ref{tab:eff_rad} indicates that interacting galaxies show modest differences in median $R_{\rm e}$ compared to their non-interacting counterparts. The median difference when all morphologies are considered together is $\sim$33 per cent, suggesting that interactions act to puff up dwarfs and increase their sizes. The dwarf ETGs show the largest discrepancy between interacting and non-interacting galaxies, with the interacting ETGs being $\sim$42 per cent larger than their non-interacting counterparts.

We proceed by comparing the trends we find in our dwarf population to those that are known in the massive galaxy regime. Studies such as \citet[][]{Shen2003}  and \citet[][]{Bernardi2014} have used SDSS data of massive ($M_{\rm \star}$ > 10$^{9.5}$ M$_\odot$) galaxies in the nearby Universe to compare galaxy sizes as a function of morphology (where they distinguish between ETGs and LTGs using a S\'ersic index threshold of 2.5). They find that the median $R_{\rm e}$ of galaxies with $M_{\rm \star}$ $\sim$ 10$^{10}$ M$_\odot$ is a factor of 2 higher in massive LTGs than in their ETG counterparts, similar to the ratio seen in our dwarf sample. However, the sizes of massive LTGs and ETGs become similar at $M_{\rm \star}$ $\sim$ 10$^{11}$ M$_\odot$, beyond which massive ETGs exhibit larger $R_{\rm e}$ than the LTGs. 

It is worth considering our findings in conjunction with the results from these studies. We note first that past studies already indicate that the slope of the effective radius -- stellar mass relation in ETGs becomes flatter as we move from the massive to the dwarf regime \citep[see e.g. Figure 4 in][]{Shen2003}. Combining our Figure \ref{fig:reff_mass} with Figure 4 in \citet{Shen2003} then suggests that the slope of this relation is indeed much shallower in dwarf ETGs than in their massive counterparts. This apparent discontinuity in the slope of the effective radius -- stellar mass relation in ETGs supports the notion \citep[see e.g.][]{Lazar2024} that the principal processes that dominate the evolution of ETGs are different in the dwarf and massive regimes. As noted already by \citet{Lazar2023} and \citet{Lazar2024}, dwarf ETGs are likely to evolve primarily via secular process like gas accretion, while massive ETGs are influenced more by interactions.

We conclude this section by considering the differences between red and blue galaxies in the massive and dwarf regimes. Recent work in the literature \citep[e.g.][]{Lange2015} suggests that in nearby ($z<0.1$) galaxies with $M_{\rm{\star}}$ $\sim$ 10$^{10}$ M$_{\odot}$ the median $R_{\rm e}$ of star forming (blue) galaxies is a factor of 2 higher than in quiescent (red) galaxies. The median values of $R_{\rm e}$ of these sub-populations become similar at $M_{\rm{\star}}$ $\sim$ 10$^{11}$ M$_{\odot}$. In the massive-galaxy regime, the relative trends between star-forming/blue and quiescent/red galaxies closely mirror those between massive ETGs and LTGs. This is largely driven by the fact that, in this regime, ETGs and LTGs dominate the red and blue populations respectively. On the other hand, our dwarf sample does not show significant differences in the median effective radius between red and blue galaxies because, as shown in \citet{Lazar2024}, more than 50 per cent of our dwarf ETGs are optically blue, in contrast with the massive regime where optically blue ETGs are rare \citep[e.g.][]{Fukugita2004,Schawinski2009,Kaviraj2007}. 

Finally, we note that, since our sample covers the redshift range $z<0.08$, in principle, it also spans a range of different physical scales. However, if we restrict our galaxies to a narrower redshift range ($0.06<z<0.08$) the trends obtained above remain unchanged. This is largely driven by the fact that $\sim$70 per cent of our galaxies actually reside in this redshift range. 


\section{Effective surface brightness}
\label{sec:sb}

Figure \ref{fig:mu_mass} presents median $\langle \mu \rangle_{\rm  e }$ vs stellar mass for our dwarf galaxies. As in Figure \ref{fig:reff_mass} above, different morphological classes are shown using different symbols, while galaxies are colour-coded using their rest-frame $(g-i)$ colour. Interacting systems are indicated using crosses. A scaling relation is present, with $\langle \mu \rangle_{\rm e}$ becoming brighter with stellar mass. Table \ref{tab:eff_sb} summarises the median values of $\langle \mu \rangle_{\rm e}$ for different galaxy populations. The dwarf ETGs are marginally brighter in this quantity that their LTG counterparts, with the featureless class being around 1 mag arcsec$^{-2}$ fainter.

Not surprisingly, with the exception of LTGs, red galaxies within each morphological class have fainter median $\langle \mu \rangle_{\rm e}$ values than their blue counterparts, with the difference being largest in the featureless galaxies. While interacting systems are larger, as described above, they have a similar median $\langle \mu \rangle_{\rm e}$ as their non-interacting counterparts. This suggests that, as already noted in \citet{Lazar2024} (see also the discussion in Section \ref{sec:colours_gradients} below), that interactions trigger star formation which boosts the surface brightness and compensates for the increase in size. It is worth noting that the featureless class is similar to ETGs in terms of $R_{\rm e}$ but differs from the ETG population in terms of $\langle \mu \rangle_{\rm e}$. Note that these results do not change if the analysis is restricted to the mass range where our featureless galaxies mostly reside (10$^{8}$ M${_\odot}$ $<$ M$_{\star}$ $<$ 10$^{8.5}$ M${_\odot}$). The spatial distributions of baryons within these classes therefore show strong differences which suggests that their formation histories are distinct. {\color{black} These findings suggest that the featureless class in low-density environments should not be considered to be a subset of the ETG population. 

In the massive-galaxy regime, around a stellar mass of $M_{\star}$ $\sim$ 10$^{10}$ M$_{\odot}$, ETGs are typically brighter in $\langle \mu \rangle_{\rm e}$ by a factor of 2.5 than LTGs \citep[see e.g.][]{Capaccioli1992,Bershady2000,Shen2003,Graham2008}. As the stellar mass approaches $M_{\star}$ $\sim$ 10$^{11.5}$ M$_{\odot}$, the median $\langle \mu \rangle_{\rm e}$ for ETGs and LTGs becomes progressively more similar. The differences between ETGs and LTGs, in terms of their effective $\langle \mu \rangle_{\rm e}$, is therefore more significant in the massive regime ($M_{\star}$ > 10$^{10}$ M${_\odot}$) than for dwarfs. We note that, while the results presented in Figures \ref{fig:reff_mass} and \ref{fig:mu_mass} are based on $i$-band images, the relative trends do not change if other filters (e.g. $g$, $r$ or $z$) are used.

It is interesting to consider how our galaxies relate to `ultra-diffuse galaxies' (UDGs), a population of faint, diffuse galaxies that exist across a variety of environments \citep[e.g.][]{Sandage1984,Impey1988,vanDokkum2015,Koda2015,Benavides2021,Marleau2021} and have sometimes been considered to be a new class of object. In Figure \ref{fig:mu_R}, we plot $\langle \mu \rangle_{\rm e}$ vs $R_{\rm e}$ for our dwarfs in the $g$, $r$ and $i$ bands, with typical UDG criteria indicated using a dashed rectangle \citep[$\langle \mu \rangle_{\rm e}$ $>$ 25 mag arcsec$^{-2}$ and $R_{\rm e}$ $>$ 1.5 kpc, see e.g.][]{Conselice2018}. 

Figure \ref{fig:mu_R} indicates that the featureless galaxies, which resemble the structure of UDGs the most, are found across a large range of surface brightnesses. Furthermore, galaxies in our sample that reside in the UDG region include members of all three morphological classes i.e. these galaxies are not morphologically distinct from the rest of the galaxy population. Taken together, this suggests that, rather than being a novel class of object, galaxies in low-density environments that satisfy the UDG criterion are simply part of the fainter and more diffuse end of the overall galaxy population \citep[see also][]{Watkins2023}. We note that a similar conclusion has been drawn about UDG-type systems in high-density environments by \citet{Conselice2018}. 

We caution, however, that, some of our galaxies in the UDG region have relatively large errors in their physical sizes and that the lower mass threshold used here ($M_{\rm{\star}}$ > 10$^{8}$ M$_{\odot}$) results in very few of our dwarfs satisfying the UDG criterion (the parameter space defined by this criterion is likely to be populated more by galaxies that have $M_{\rm{\star}}$ < 10$^{8}$ M$_{\odot}$ and are fainter). Probing galaxies fainter than those in this study is likely needed to consolidate this result.


\section{Colour profiles and gradients}
\label{sec:colours_gradients}

In Figure \ref{fig:col_grad_prof}, we present median $(g-i)$ colour profiles and colour gradients in our dwarf galaxies. Note that, throughout this section, to ensure that our results are not noisy, we only calculate the colour in a given radial bin if it has at least 10 data points. The top row presents these quantities for the full mass range spanned by our sample (10$^{8}$ M$_\odot$ > $M_{\rm \star}$ > 10$^{9.5}$ M$_\odot$), while the middle and bottom rows correspond to the lower and upper halves of our mass range respectively. The left and right-hand columns show the $(g-i)$ colour and its gradient as a function of the radius normalised by $R_{\rm e}$. As shown in \citet{Lazar2024}, the featureless galaxies only appear in the lower half of our stellar mass range, as result of which this morphological class is missing in the bottom row. In each panel the solid line represents the running median value, while the shaded region indicates the error in the running median.

\begin{figure}
 \centering
 \includegraphics[width=0.42\textwidth]{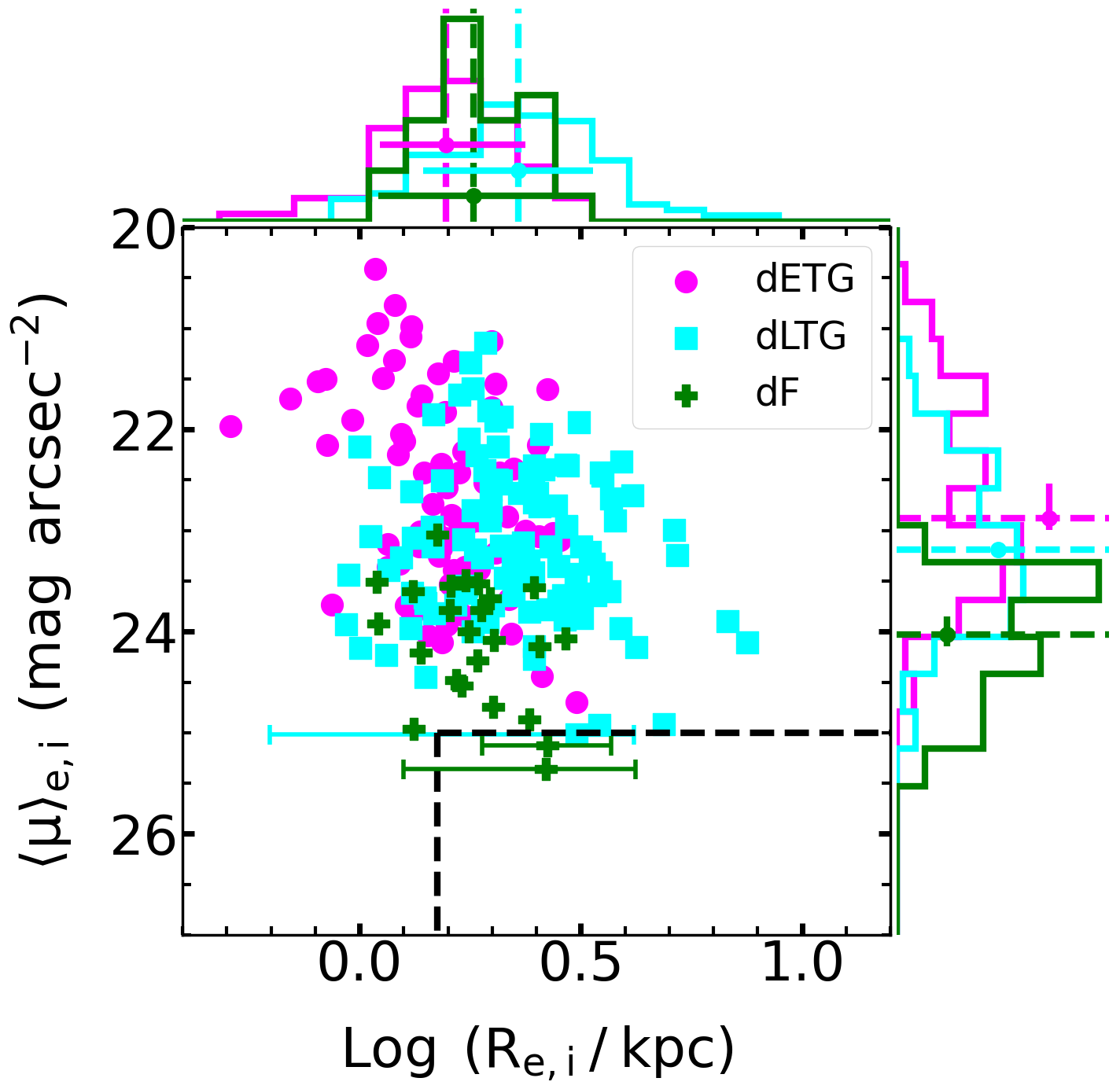}
 \includegraphics[width=0.42\textwidth]{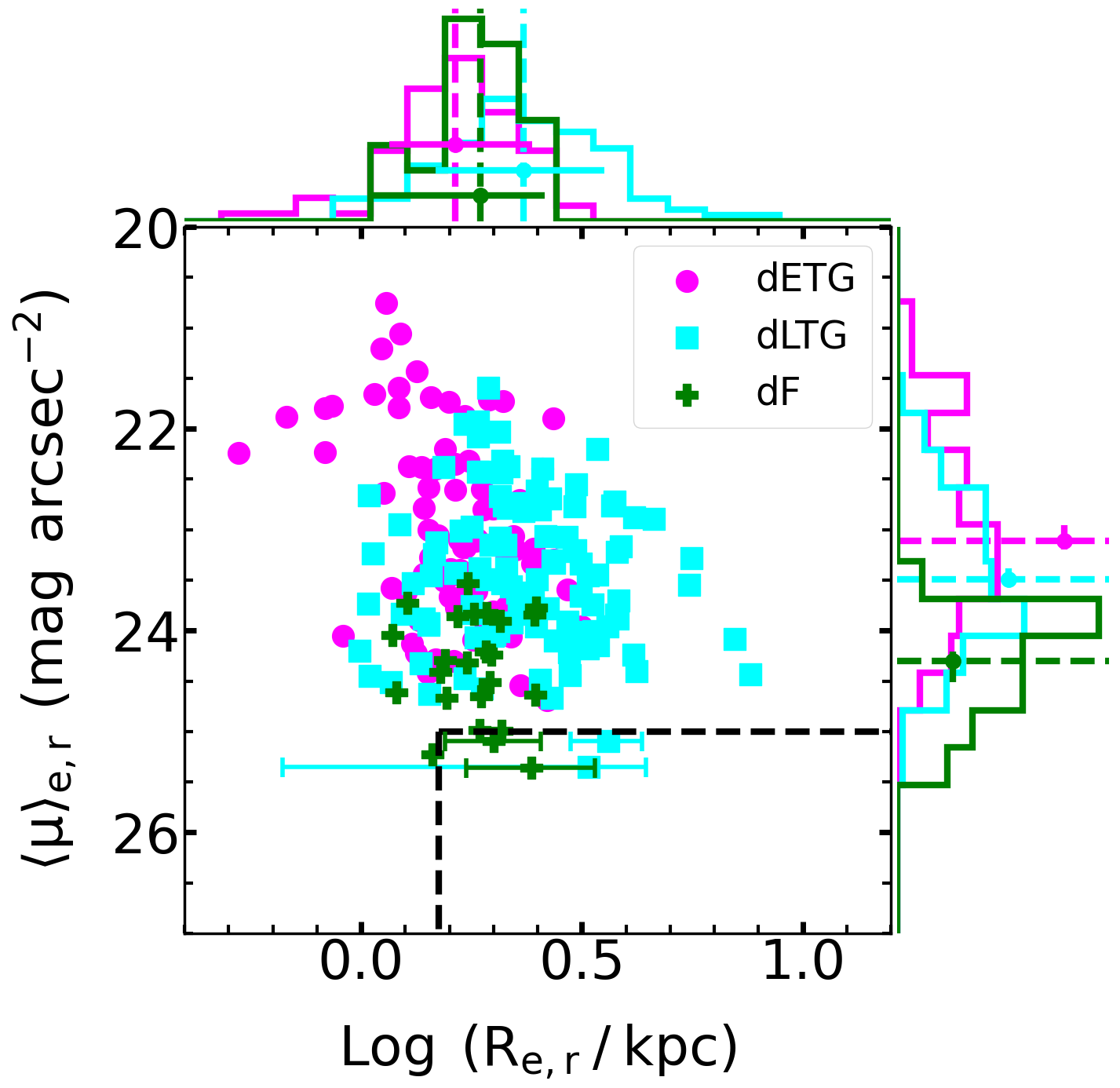}
 \includegraphics[width=0.42\textwidth]{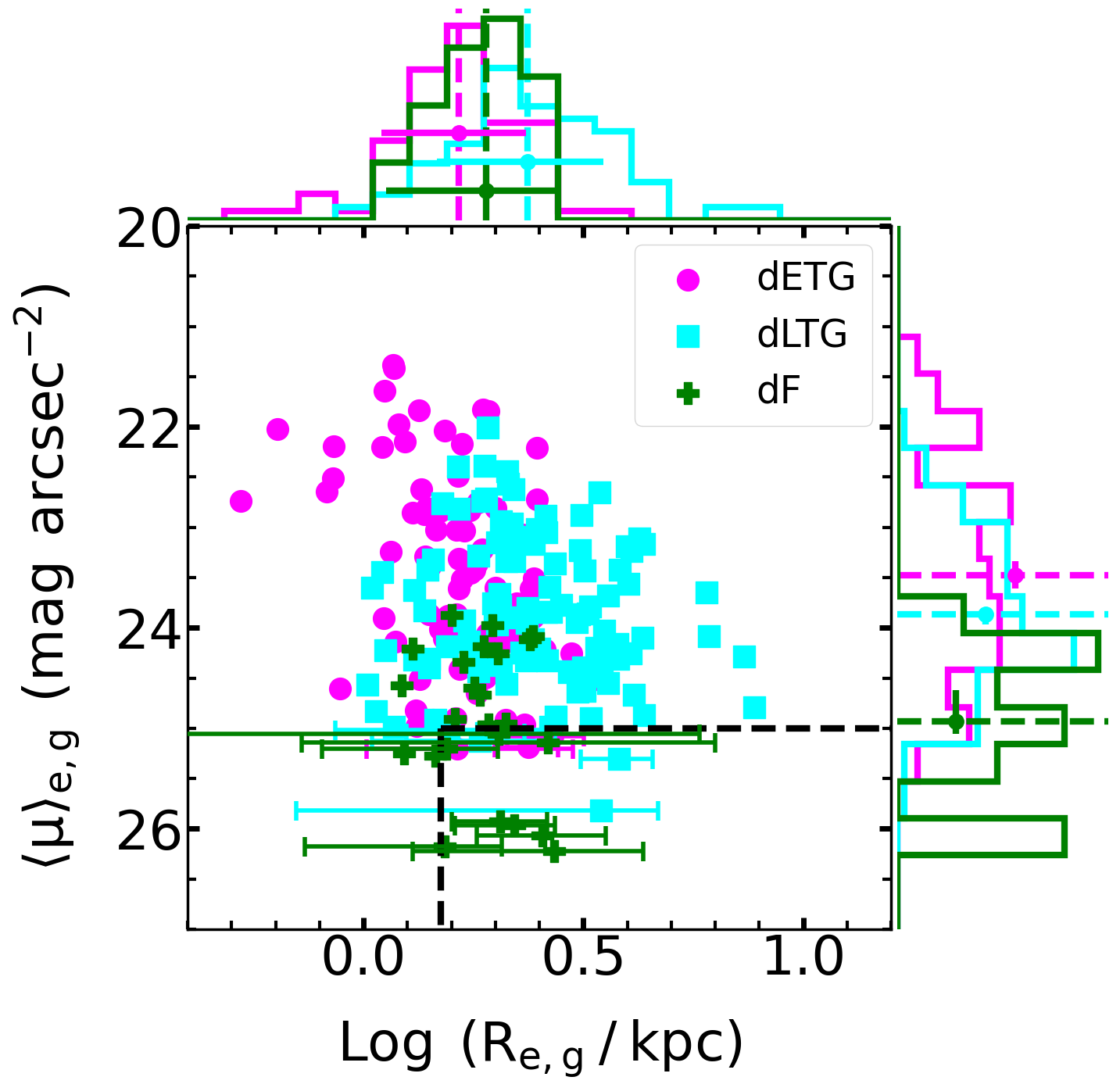}
 \caption{$\langle \mu \rangle_{\rm e}$ as a function of $R_{\rm e}$ for our dwarf galaxies. The top, middle and bottom panels correspond to measurements in the $i$, $r$ and $g$ bands respectively. Different morphological classes are shown using different colours. The dashed lines within the panel correspond to UDG selection criteria \citep[e.g.][]{Conselice2018}. Error bars of galaxies that fall in the UDG region are shown (and omitted for other galaxies for clarity). The dashed lines on the histograms represent median values, with the uncertainties on the medians calculated via bootstrapping.}
 \label{fig:mu_R}
\end{figure}

\begin{figure*}
 \centering
 \includegraphics[width=0.44\textwidth]{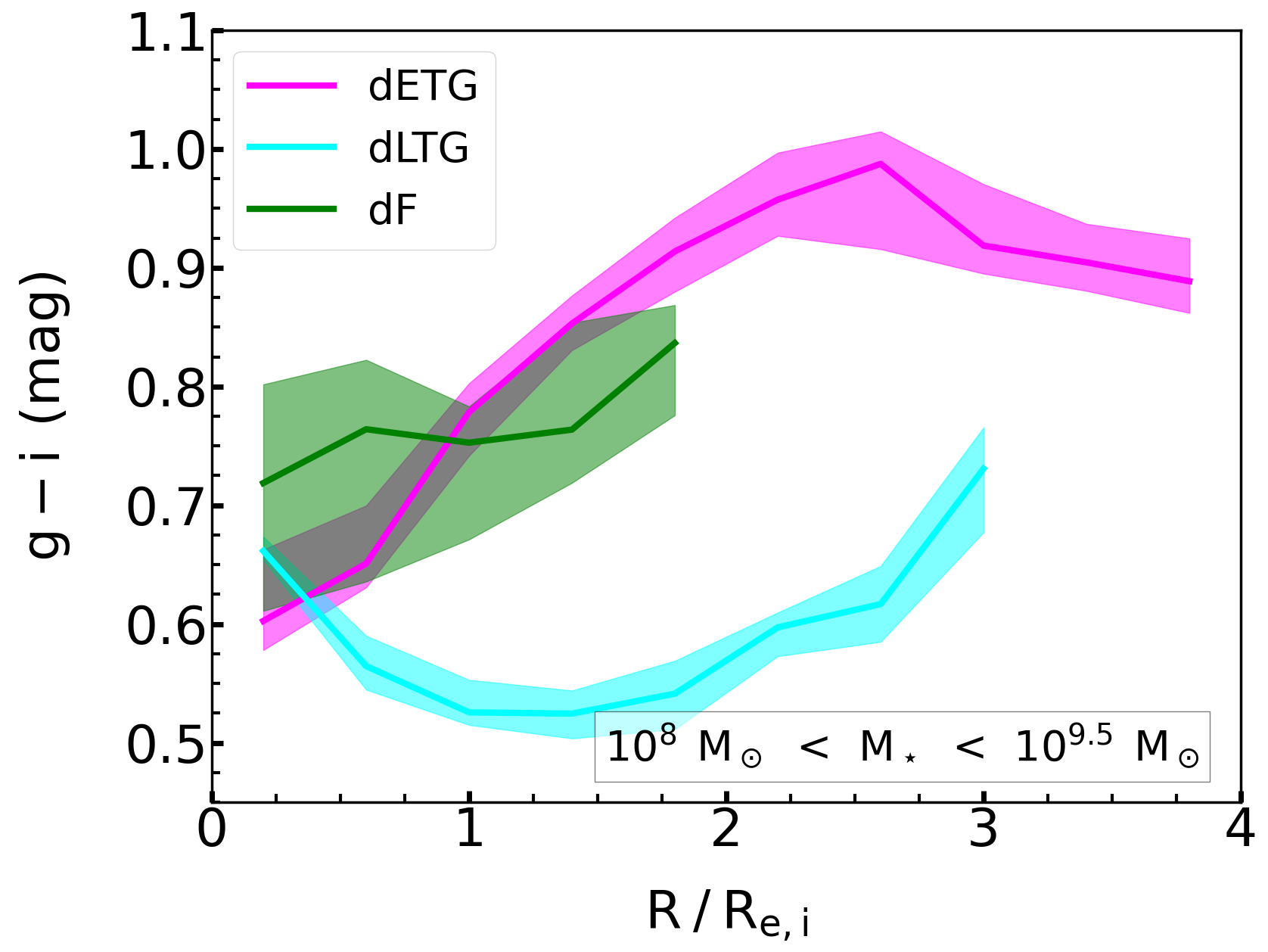}
 \includegraphics[width=0.455\textwidth]{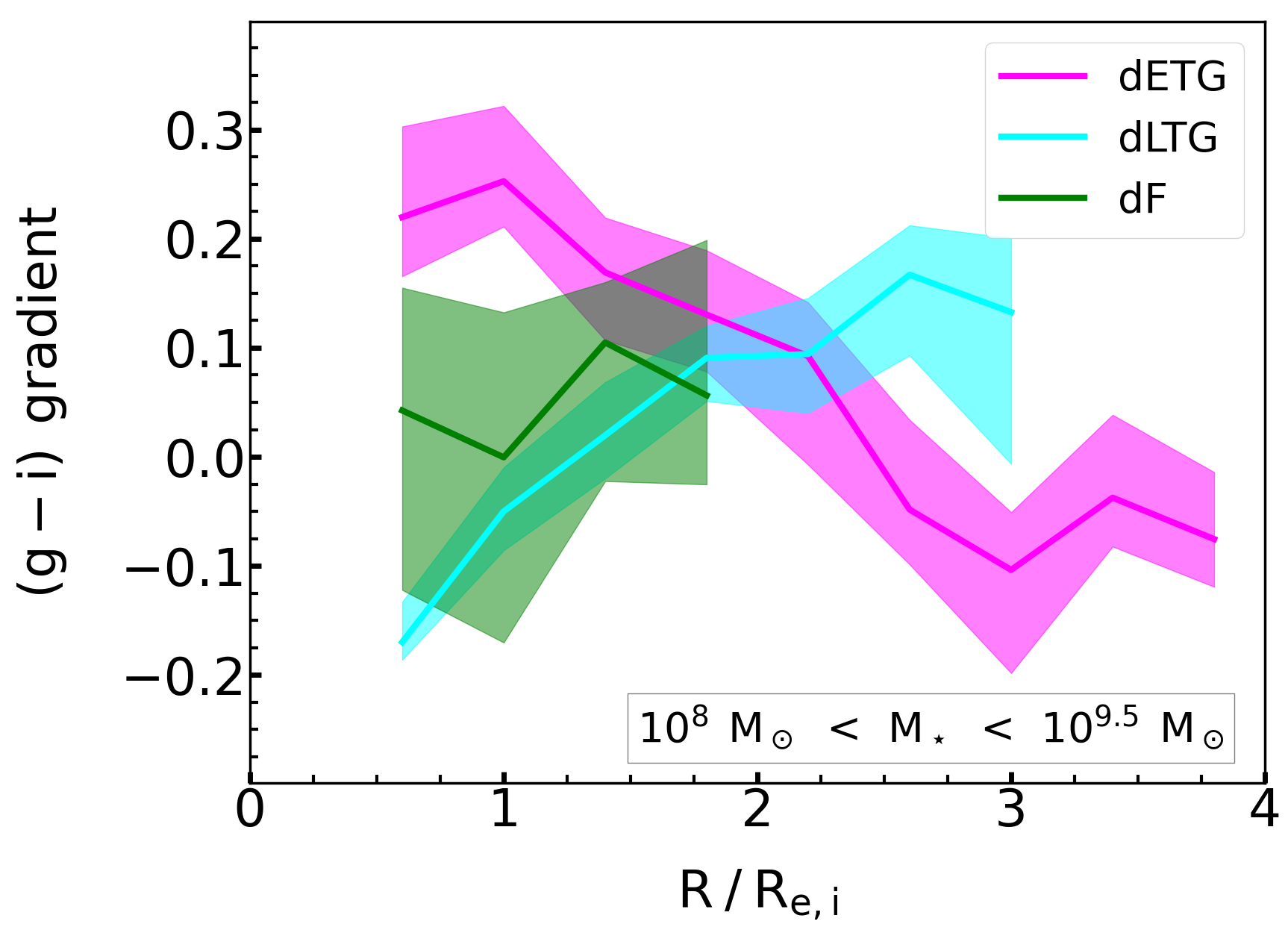}
 \includegraphics[width=0.44\textwidth]{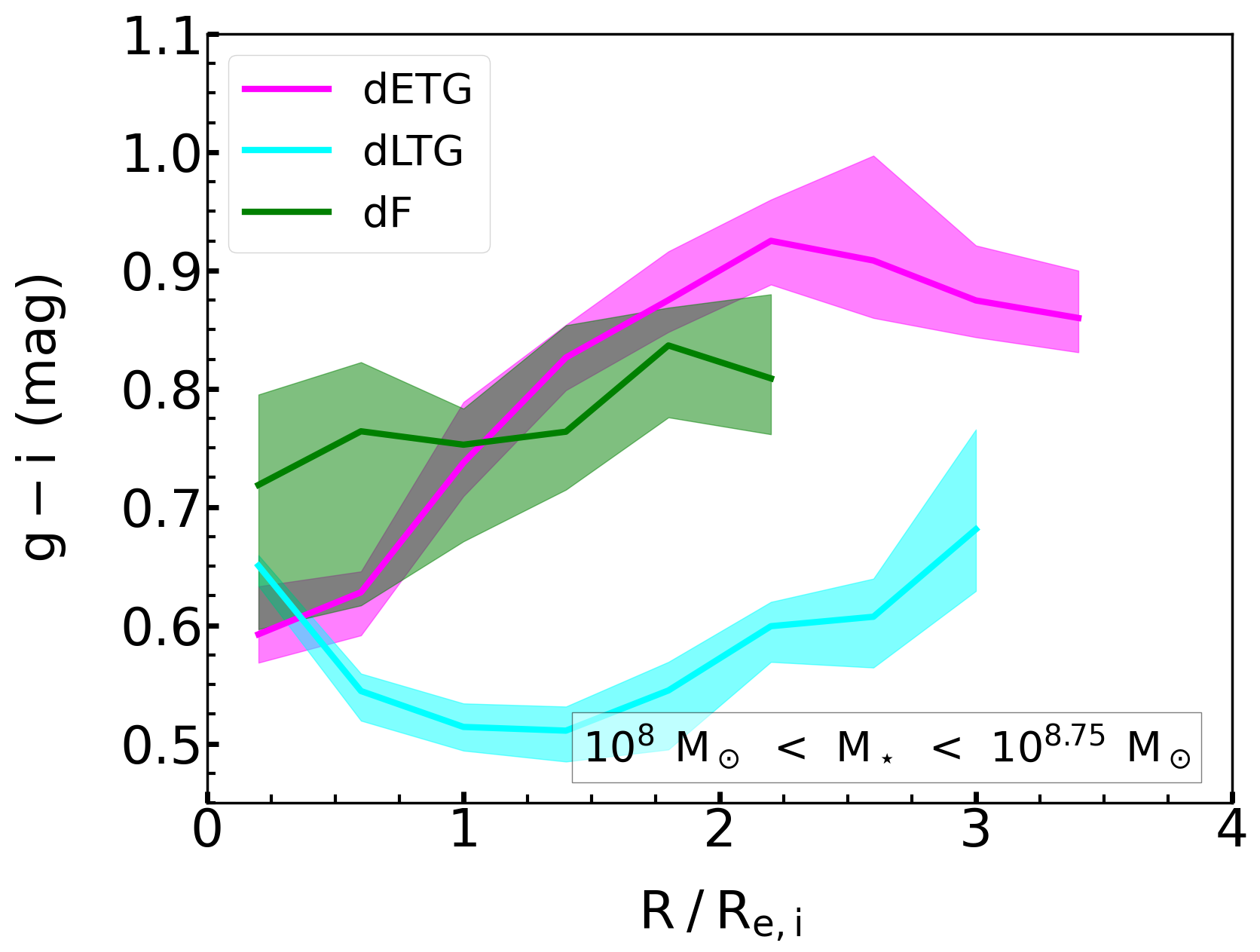}
 \includegraphics[width=0.455\textwidth]{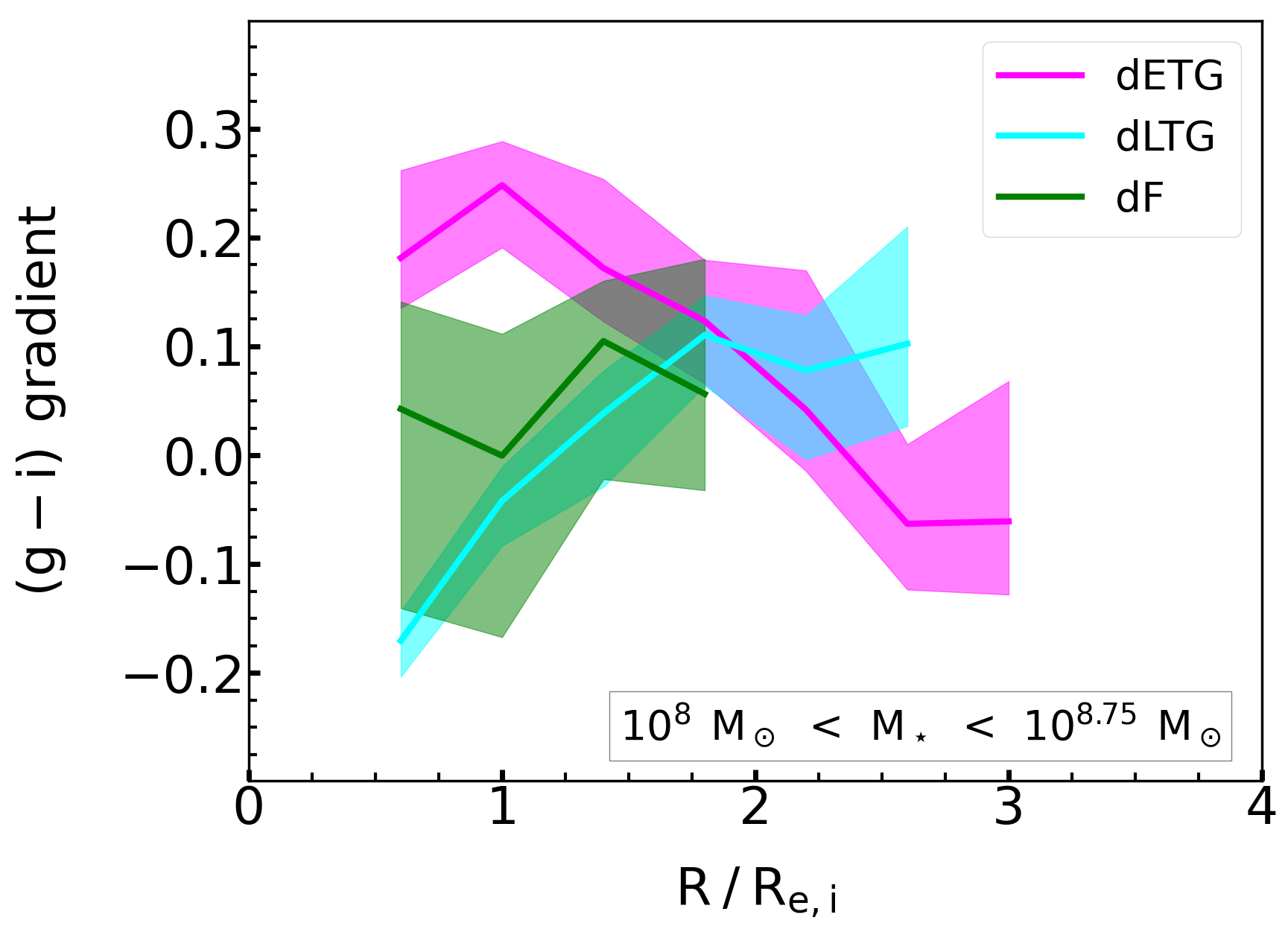}
 \includegraphics[width=0.44\textwidth]{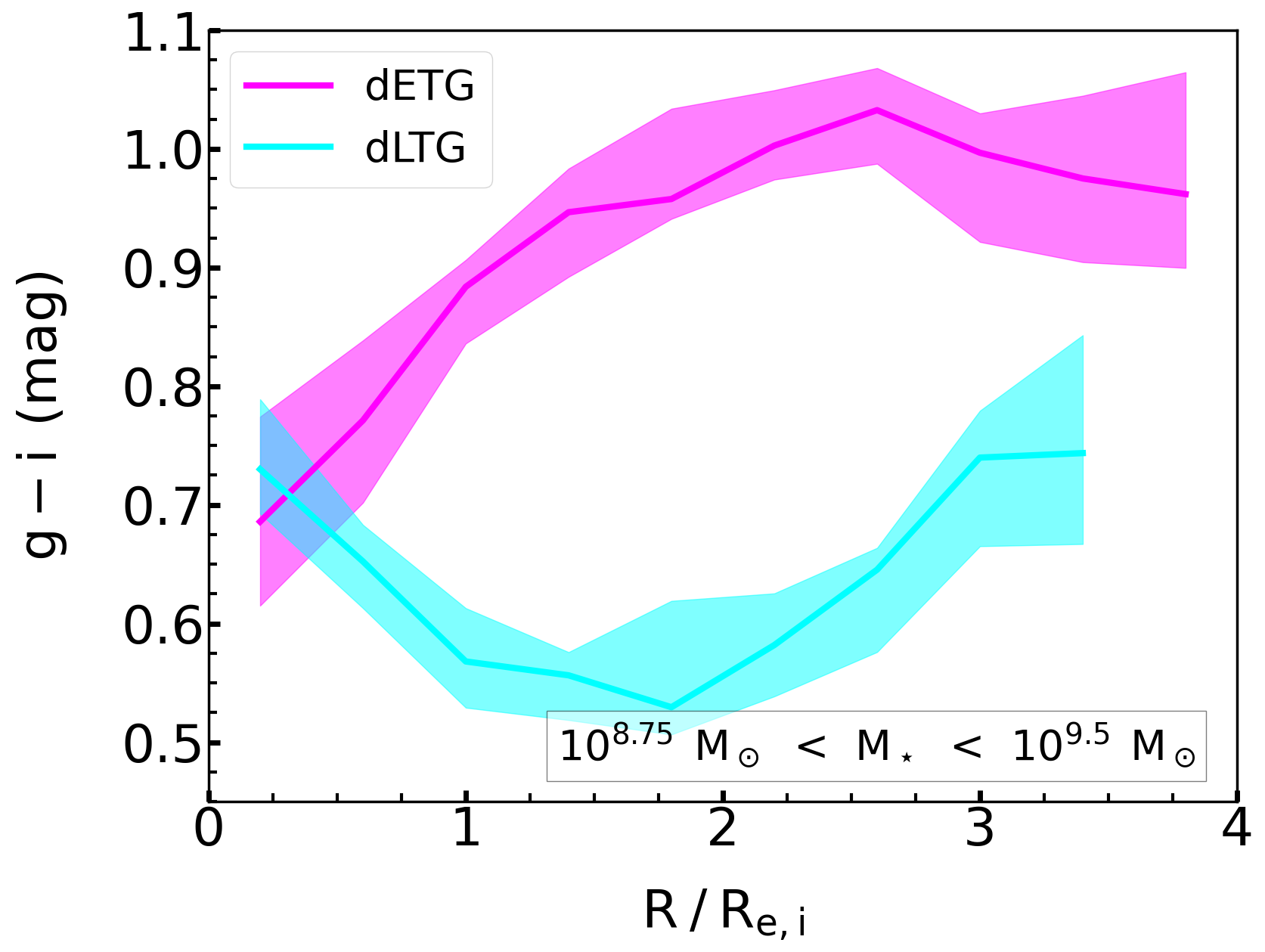}
 \includegraphics[width=0.455\textwidth]{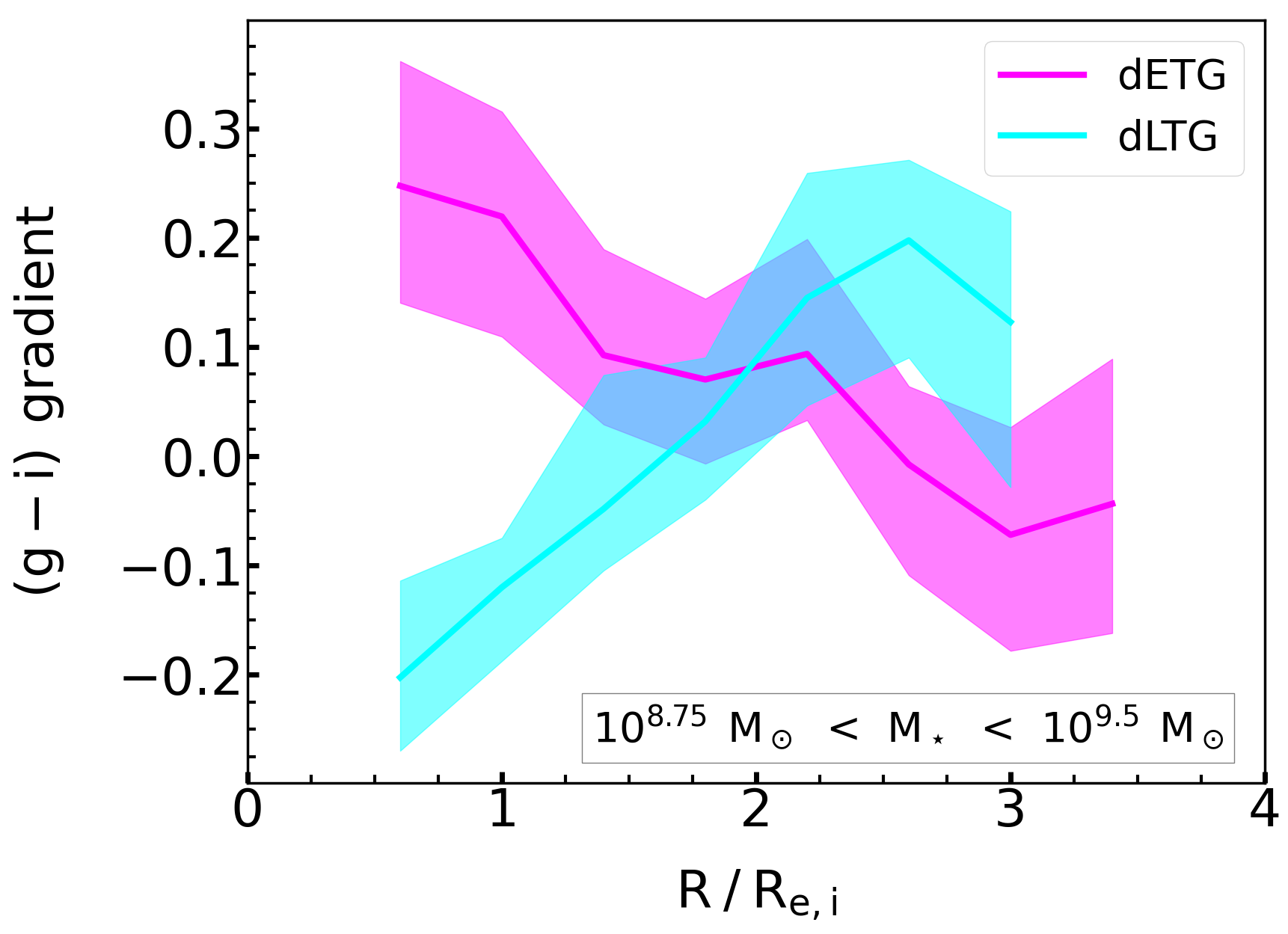}
 \caption{Median $(g-i)$ colour profiles (left-hand column) and colour gradients (right-hand column) in our dwarf galaxies. The top row presents these quantities for the full mass range spanned by our sample (10$^{8}$ M$_\odot$ > $M_{\rm \star}$ > 10$^{9.5}$ M$_\odot$), while the middle and bottom rows correspond to the lower and upper halves of our mass range respectively. The left and right-hand columns show the $(g-i)$ colour and its gradient, as a function of the radius normalised by $R_{\rm e}$, respectively. As shown in \citet{Lazar2024}, all featureless galaxies occupy the lower half of our stellar mass range, as a result of which this morphological class is missing in the bottom row. In each panel the solid line represents the running median and the shaded region indicates the bootstrapped uncertainty on the median. The gradient is calculated by dividing the difference of the colour values of adjacent radial bins. The uncertainties shown using the shaded regions are calculated using standard error propagation.} 
 \label{fig:col_grad_prof}
\end{figure*}

We first consider dwarfs across our full stellar mass range. Dwarf LTGs exhibit a `U' shaped profile with a negative gradient in their central regions out to $\sim$1.5 $R_{\rm e}$. At this point the $(g-i)$ colour reaches a plateau, which represents the bluest region of the galaxy. Beyond $\sim$2 $R_{\rm e}$ the colour progressively reddens and the gradient becomes positive. On the other hand, the dwarf ETGs exhibit positive colour gradients until $\sim$2.5 $R_{\rm e}$, beyond which the colour gradient becomes negative and the colours in the outskirts of the dwarf ETGs become progressively bluer. It is worth noting that even though their colour gradients are the opposite of each other, dwarf ETGs and LTGs show similar colours in their central regions (i.e. at < 0.5 $R_{\rm e}$). Dwarf featureless galaxies show a shallow positive (i.e. close to flat) gradient throughout their colour profiles. They generally exhibit redder central regions (out to around $R$ $\sim$ $R\rm _{e}$) than both ETGs and LTGs, suggesting the existence of comparatively older stellar populations. At larger radii ($R > R_{\rm e}$) the featureless galaxies are slightly bluer than ETGs but significantly redder than their LTG counterparts.

\begin{figure*}
 \centering
 \includegraphics[width=0.45\textwidth]{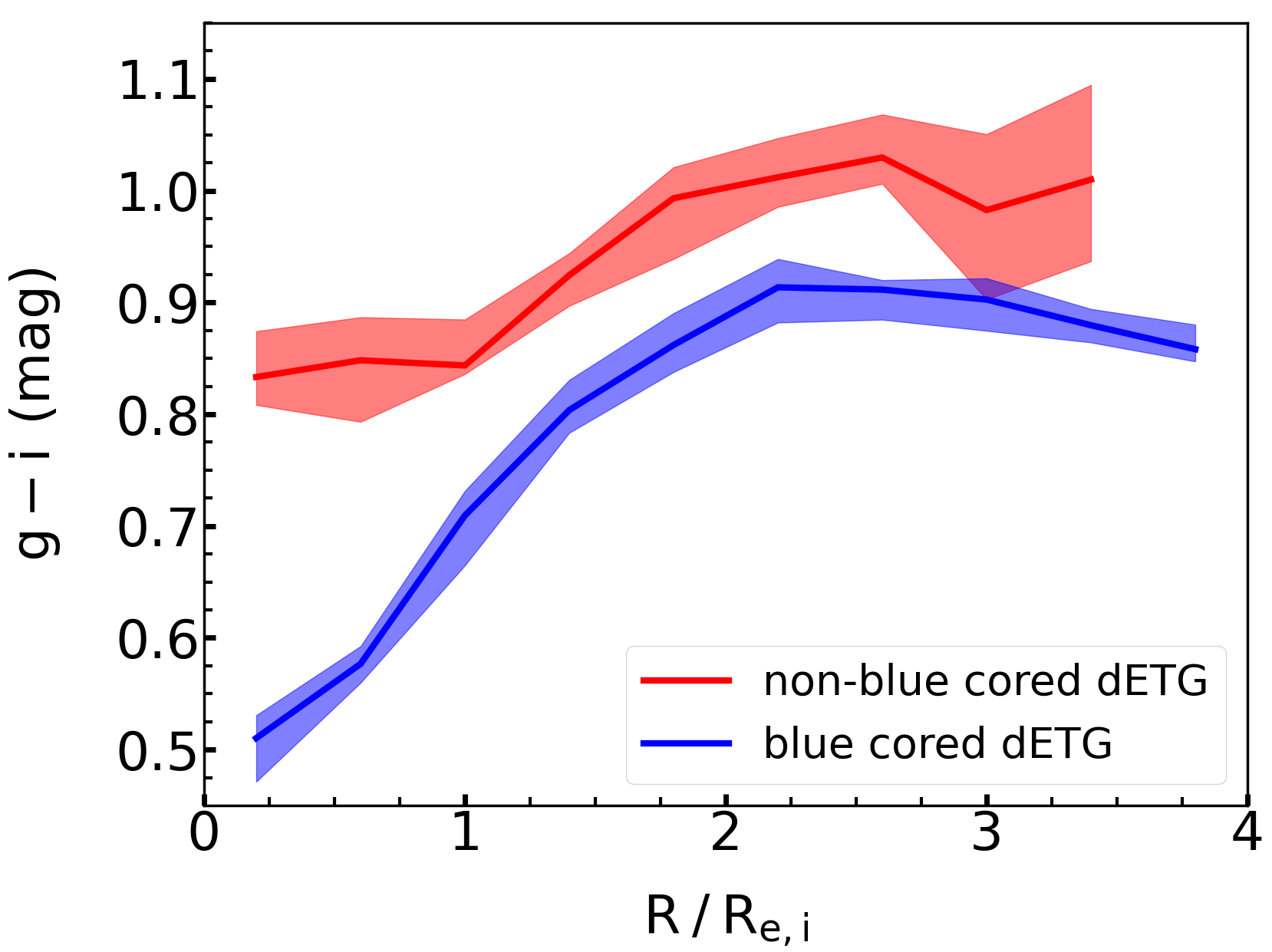}
 \includegraphics[width=0.47\textwidth]{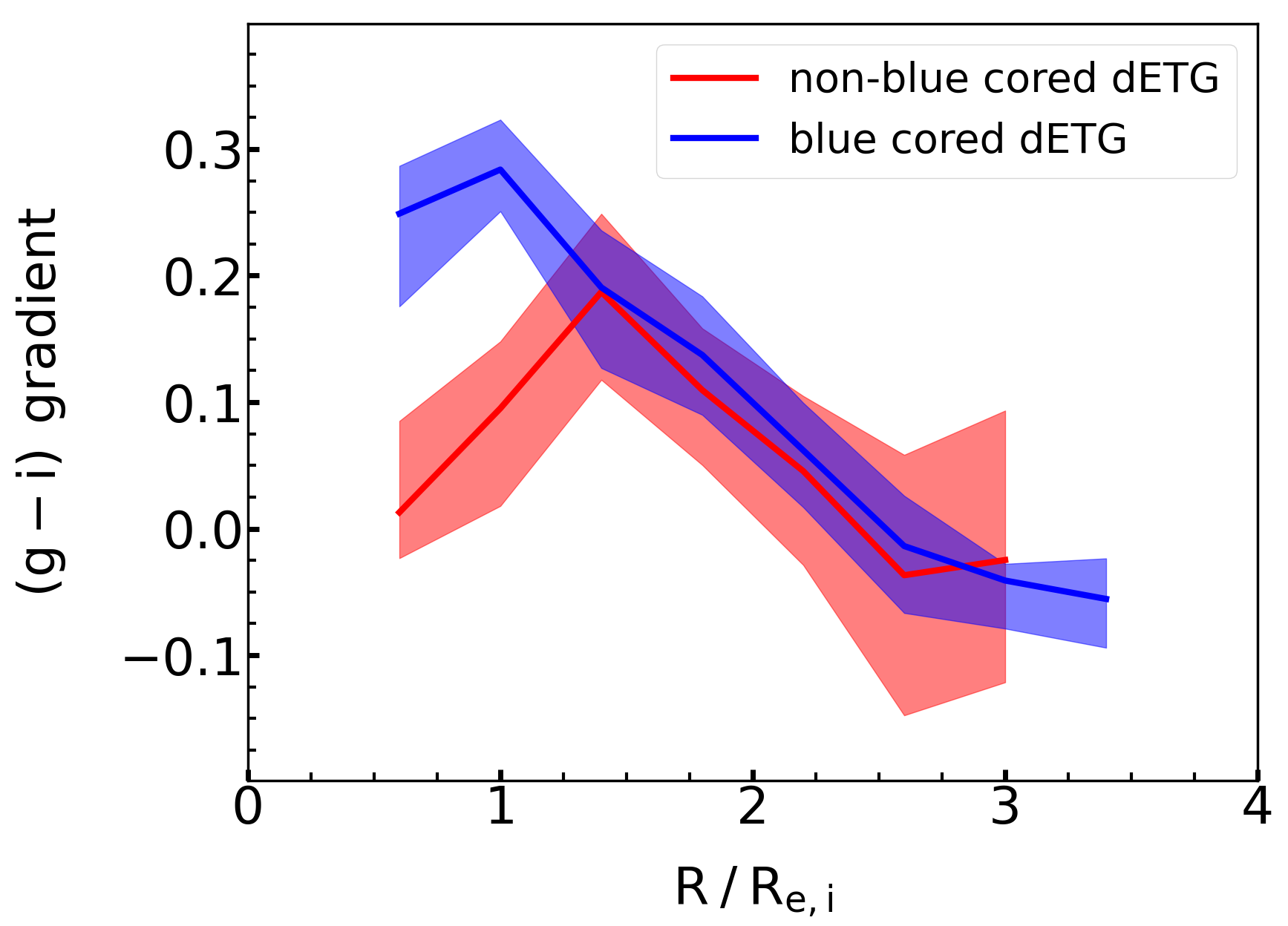}
 \caption{Median $(g-i)$ colour profiles (left-hand panel) and colour gradients (right-hand panel) in blue-cored and non blue-cored dwarf ETGs. In each panel the solid line represents the running median, while the shaded region indicates its uncertainty.}
 \label{fig:bl_core}
\end{figure*}

The middle and bottom rows of Figure \ref{fig:col_grad_prof} show that there are no qualitative differences between the two mass bins in their colour profiles and gradients. However, we note that both the ETGs and LTGs have redder central regions (by $\sim$0.1 mag at $R<R_{\rm e}$) in the high mass bin compared to their counterparts in the low mass bin. Dwarf ETGs in the high mass bin also exhibit redder colours at all radii than ETGs in the low mass bin. 

We proceed by comparing our findings to what is known in the massive-galaxy regime. The majority of massive LTGs have negative colour gradients at $R$ < $R_{\rm e}$ which flatten or turn positive at outer radii \citep[e.g.][]{Moth2002,Taylor2005,Bakos2008,Liu2009,Tortora2010,DSouza2014,Watkins2019,Liao2023}. This behaviour resembles that seen in our dwarf LTGs. The physical processes that cause the reddening of the colour in the outskirts are still a matter of debate but some models postulate that this behaviour could be driven by the outward radial migration of evolved stars via intrinsic secular processes \citep[e.g. through spirals or bars][]{Debattista2006,Roskar2008,Munoz2013} combined with a star formation cut-off at a certain surface density threshold in the outer regions of the galaxy \citep[e.g.][]{Martin2001}.

Massive ETGs ($M_{\rm\star}$ > 10$^{9.5}$ M$_\odot$) typically have flat or negative colour gradients \citep[e.g.][]{Kormendy1989,Peletier1990,DePropris2005,Marian2018}. 
These gradients and the resultant bluer colours in the outskirts of these systems are thought to be caused by satellite accretion events \citep[e.g.][]{Kaviraj2009,LaBarbera2012,Ferreras2017,Martin2018_sph,Jackson2023}, with the stars from the satellites, which are typically bluer, settling in the outskirts of the massive ETGs. It is worth noting, however, that at the lower mass end of the massive regime (10$^{9.5}$ M$_\odot$ < $M_{\rm\star}$ < 10$^{10.5}$ M$_\odot$), a minority (between 10 and 30 per cent) of ETGs in relatively low-density environments do show positive gradients \citep[e.g.][]{Tamura2000,Ferreras2005,Jiang2011,Suh2010,Tortora2010}, driven by recent star formation activity in their cores within the last $\sim$1 Gyr. When considered together with our results, this suggests that the incidence of blue ETGs (driven by blue central regions) does indeed increase with decreasing stellar mass.

The differences in the colour gradients between massive and dwarf ETGs suggests that the principal mode of stellar mass growth is likely to be different. The evolution of massive ETGs is likely to be driven by `inside-out' growth, where the core of the galaxy forms earlier in cosmic time and material is then accreted through minor mergers in the outskirts. However, ETGs in the dwarf regime appear to experience `outside-in' growth, which is likely to be driven less by local environment and interactions and more by stellar feedback and secular processes \citep[e.g.][]{Pipino2004,Zhang2012,Perez2013,Cheng2020,Ge2024}. 

This `outside-in' scenario is thought to proceed via star formation starting in the outskirts of the galaxy (e.g. via gas accretion) and propagating inwards causing gas heating and expansion due to supernova events. Since low mass galaxies have shallow potential wells, some of the gas is blown away from the outer regions of the galaxy. The gas that remains cools down and sinks deeper towards the galactic centre giving rise to further star formation events. As a result, the galaxy is left with older stellar populations in the outer regions (since the gas reservoir is depleted due to the supernova-driven galactic wind) and younger stellar populations in the galactic centre. Note that our results are consistent with the conclusions of past studies \citep[e.g.][]{Pan2015} that have suggested that the transition between inside-out and outside-in growth takes place somewhere between $M_{\rm \star}$ $\sim$ 10$^{10}$ M$_\odot$ and $M_{\rm \star}$ $\sim$ 10$^{10.5}$ M$_\odot$.

It is worth separately exploring the colour profiles of two interesting sub-populations of dwarfs: ETGs which have visually identified blue cores and dwarfs that have been flagged as interacting in \citet{Lazar2024}. As noted in that study, significant blue cores appear to be common in dwarf ETGs, with around 46 per cent of these systems showing cores that are clearly visible by eye (two such examples can be seen in the top row of Figure \ref{fig:images}). Figure \ref{fig:bl_core} shows that the blue cored dwarf ETGs are bluer than their non-cored counterparts at all radii. However, beyond $R \sim$ 1.5 $R_{\rm e}$, the colour gradients become indistinguishable in the cored and non-cored populations (i.e. the colour profiles are roughly parallel to each other). At $R$ < 1.5 $R_{\rm e}$, however, the colour of the blue-cored ETGs diverges significantly from their non-cored counterparts, reaching an offset of -0.35 mag in the very centre. Most of the light from the blue core therefore appears to originate from the region enclosed within $\sim$ 1.5 $R_{\rm e}$ in the blue-cored systems. 

\begin{figure}
 \centering
 \includegraphics[width=0.45\textwidth]{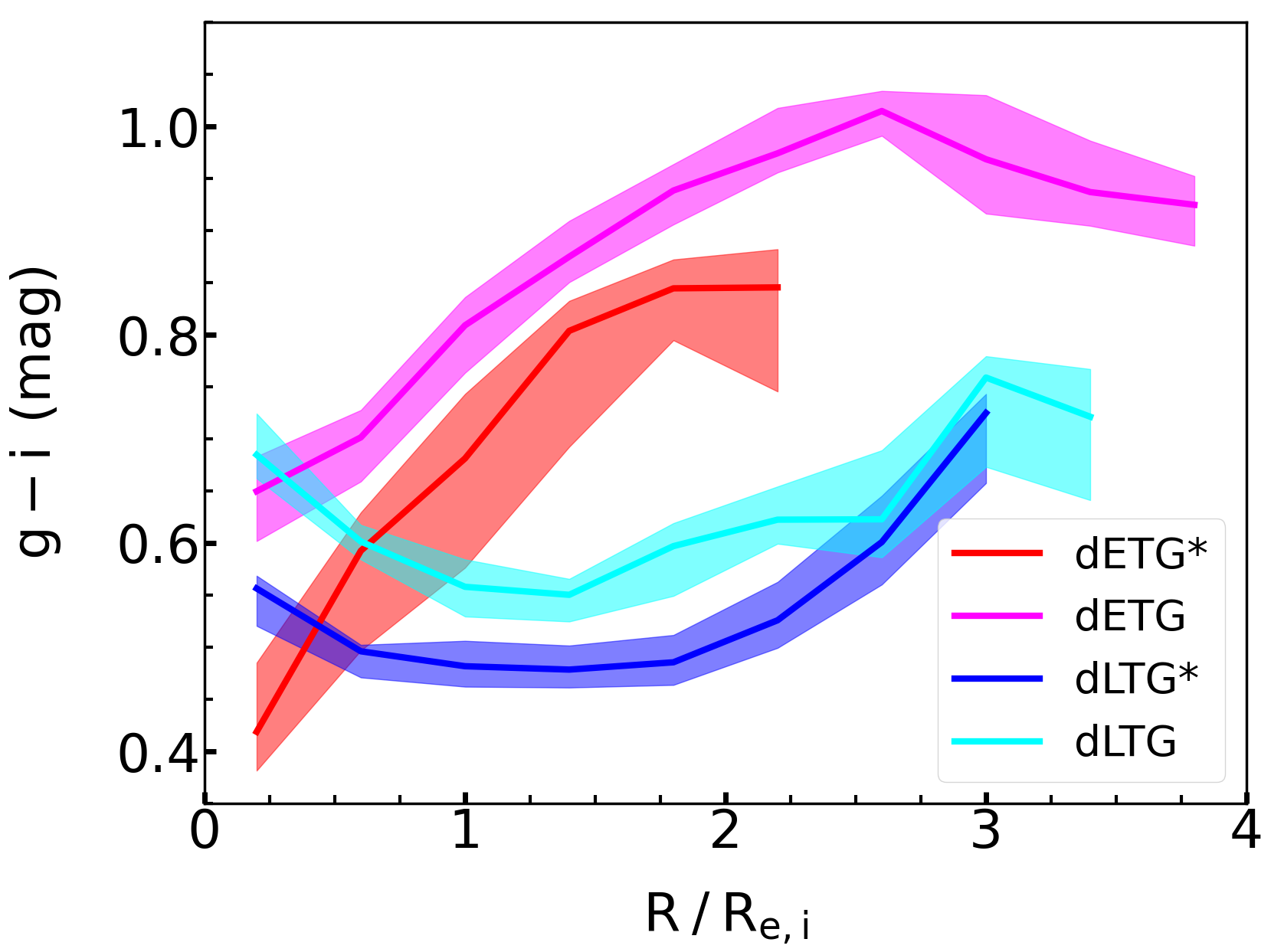}
 \caption{Median $(g-i)$ colour profiles in interacting (shown using a `*') and non-interacting dwarf ETGs and LTGs. In each panel the solid line represents the running median, while the shaded region indicates its uncertainty.}
 \label{fig:tidal}
\end{figure}

We conclude our study by exploring dwarfs that have been flagged as interacting in the visual inspection performed in \citet{Lazar2024}. Figure \ref{fig:tidal} shows that, regardless of morphology, interacting dwarfs tend to be bluer than their non-interacting counterparts at all radii. This is consistent with the finding in \citet{Lazar2024} that the median integrated colour of interacting dwarfs is typically bluer than their non-interacting counterparts. The bluer colours suggest that the interactions are enhancing the star formation activity in these systems, as has been suggested in the recent literature \citep[e.g.][]{Stierwalt2015,Martin2021}. In particular, the consistent offset between the colour profiles of the interacting and non-interacting dwarfs suggests that star formation may be enhanced across the entire extent of the galaxy (rather than being localised in, say, the central regions). It is worth noting that the blueward colour offset in the interacting systems is largest in the very central regions ($R$ $<$ 0.5 $R_{\rm e}$) which suggests that the star formation enhancement is largest in the galactic centre. These results appear consistent with recent studies \citep[e.g.][]{Privon2017} which show that star formation enhancement via interactions in dwarfs could indeed be spatially extended and triggered by large-scale tidal compressions in the inter-stellar medium, which act across the inner and outer regions of the galaxy. Note that the apparent truncation in the $(g-i)$ colour for interacting dwarf ETGs is artificial and caused by the fact that the number of data points for $R$ $>$ 2 $R_{\rm e,i}$ drops below 10 (the threshold we apply for calculating the median colour in a given radial bin).


\section{Summary}
\label{sec:summary}

We have used a complete, unbiased sample of 211 nearby ($z<0.08$) dwarf (10$^{8}$ M$_{\odot}$ < M$_{\rm{\star}}$ < 10$^{9.5}$ M$_{\odot}$) galaxies, to study the structure of nearby dwarf galaxies in low-density environments. In particular, we have studied galaxy size (parametrised by $R_{\rm e}$), $\langle \mu \rangle_{\rm e}$ and the $(g-i)$ colour profile and colour gradient, as a function of stellar mass and morphology. Our main conclusions are as follows.

\begin{itemize}

    \item Dwarf ETGs and LTGs are well separated in $R_{\rm e }$, with the median $R_{\rm e }$ of LTGs being around a factor of 2 larger than that of the ETGs. The featureless dwarfs lie in between the ETGs and LTGs, with a median $R_{\rm e }$ that is around a factor of 1.2 larger than their ETG counterparts. 
    
    \item The median $R_{\rm e }$ of red (i.e. quenched) and blue (i.e. star-forming) galaxies do not show significant differences within each dwarf morphological class. When all morphological classes are considered together, there is a weak trend of red galaxies being more compact than blue galaxies. 

    \item Dwarf ETGs are marginally brighter in median $\langle \mu \rangle_{\rm e}$ that their LTG counterparts, with the featureless class being around 1 mag arcsec$^{-2}$ fainter.

     \item The colour profiles and gradients of dwarf ETGs differ significantly from their massive counterparts. Dwarf ETGs typically show positive gradients (i.e. bluer central regions), while massive ETGs either have red colours throughout or a red core with blue outer regions (i.e. negative or flat gradients). The divergence in colour profiles suggests that ETGs have different formation channels in the dwarf and massive regimes. Massive ETGs are likely to evolve `inside-out', as a result of minor mergers adding stars to their outskirts. Dwarf ETGs, on the other hand, evolve `outside-in', as a result of stellar-feedback driven galactic winds being more effective at quenching their outskirts while star formation continues in their central regions. 

    \item Dwarf LTGs show similar colour profiles as their massive counterparts, exhibiting negative colour gradients in their central regions ($R$ $<$ 1.5 $R_{\rm e}$) and positive gradients in the outer regions ($R$ $>$ 2 $R_{\rm e}$).  

    \item Interacting systems are larger but have a similar median $\langle \mu \rangle_{\rm e}$ as their non-interacting counterparts. This suggests that interactions trigger star formation, which boosts the surface brightness and compensates for the increase in size. Regardless of morphology, the colour profiles of interacting dwarfs are bluer than their non-interacting counterparts at all radii, with the blueward colour offset being largest in the very central regions. This indicates that, in the dwarf regime, the enhancement of star formation due to interactions typically takes place across the entire extent of the galaxy. 

    \item The dwarf featureless and ETG classes do not differ significantly in terms of their median $R_{\rm e }$ but do differ in their median $\langle \mu \rangle_{\rm e}$. The distributions of baryons within these classes therefore show strong differences, suggesting that their formation histories are different. This suggests that, in low-density environments, the featureless class should not be considered to be a subset of the ETG population. We note, however, that our sample of dwarf featureless galaxies is relatively small due to our lower stellar mass limit of 10$^8$ M$_\odot$. Larger samples of galaxies extending to lower stellar masses are likely needed to confirm this result. 

    \item Dwarfs in our sample that reside in the UDG  region in the $R_{\rm e}$ vs $\langle \mu \rangle_{\rm e}$ parameter space include members of all morphological classes and are a continuous extension of the galaxy population towards lower values of $\langle \mu \rangle_{\rm e}$. This suggests that, rather than being a novel class of object, galaxies in low-density environments that satisfy the UDG criterion are simply part of the fainter and more diffuse end of the overall galaxy population. 

    \item The prominent blue cores that are visually identified in around 46 per cent of dwarf ETGs extend out to $\sim$ 1.5 $R_{\rm e}$.

\end{itemize}


\section*{Acknowledgements}

We warmly thank the anonymous referee for several constructive suggestions which helped us improve the original manuscript. We thank Pierre-Alain Duc, Elizabeth Sola and Liza Sazonova for many interesting discussions. IL and BB acknowledge PhD studentships from the Centre for Astrophysics Research at the University of Hertfordshire. SK and IL acknowledge support from the STFC (grant number ST/Y001257/1). SK and AEW acknowledge support from the STFC (grant number ST/X001318/1). SK also acknowledges a Senior Research Fellowship from Worcester College Oxford. For the purpose of open access, the authors have applied a Creative Commons Attribution (CC BY) licence to any Author Accepted Manuscript version arising from this submission.


\section*{Data Availability}

{\color{black}The structural parameters produced in this work are available via common online repositories. They can also be obtained by contacting the authors.}


\bibliographystyle{mnras}
\bibliography{paper}







\bsp	
\label{lastpage}
\end{document}